\documentclass{IEEEtran}
\usepackage{amsmath,amsfonts}
\usepackage{algorithmic}
\usepackage{algorithm}
\usepackage{array}
\usepackage[caption=false,font=normalsize,labelfont=sf,textfont=sf]{subfig}
\usepackage{textcomp}
\usepackage{stfloats}
\usepackage{url}
\usepackage{enumitem}
\usepackage{amsthm}
\usepackage{amssymb}
\usepackage{cite}
\usepackage{xcolor}
\usepackage[nolist]{acronym}
\usepackage{graphicx,scalerel}

\begin{document}
\begin{acronym}

\acro{2D}{Two Dimensions}%
\acro{2G}{Second Generation}%
\acro{3D}{Three Dimensions}%
\acro{3G}{Third Generation}%
\acro{3GPP}{Third Generation Partnership Project}%
\acro{3GPP2}{Third Generation Partnership Project 2}%
\acro{4G}{Fourth Generation}%
\acro{5G}{Fifth Generation}%

\acro{AI}{Artificial Intelligence}%
\acro{AoA}{Angle of Arrival}%
\acro{AoD}{Angle of Departure}%
\acro{AR}{Augmented Reality}%
\acro{AP}{Access Point}
\acro{AE}{Antenna Element}
\acro{AC}{Anechoic Chamber}
\acro{AUT}{Antenna Under Test}
\acro{AD}{Anderson-Darling}%

\acro{BER}{Bit Error Rate}%
\acro{BPSK}{Binary Phase-Shift Keying}%
\acro{BRDF}{ Bidirectional Reflectance Distribution Function}%
\acro{BS}{Base Station}%

\acro{CA}{Carrier Aggregation}%
\acro{CDF}{Cumulative Distribution Function}%
\acro{CDM}{Code Division Multiplexing}%
\acro{CDMA}{Code Division Multiple Access}%
\acro{CPU} {Central Processing Unit}
\acro{CUDA}{Compute Unified Device Architecture}
\acro{CDF}{Cumulative Distribution Function}
\acro{CI}{Confidence Interval}
\acro{CVRP}{Constrained-View Radiated Power}
\acro{CATR}{Compact Antenna Test Range}
\acro{CV}{Coefficient of Variation}
\acro{CTIA}{Cellular Telephone Industries Association}
 
\acro{D2D}{Device-to-Device}%
\acro{DL}{Down Link}%
\acro{DS}{Delay Spread}%
\acro{DAS}{Distributed Antenna System}
\acro{DKED}{double knife-edge diffraction}
\acro{DUT}{Device Under Test}
\acro{DR}{Dynamic Range}


\acro{EDGE}{Enhanced Data rates for GSM Evolution}%
\acro{EIRP}{Equivalent Isotropic Radiated Power}%
\acro{eMBB}{Enhanced Mobile Broadband}%
\acro{eNodeB}{evolved Node B}%
\acro{ETSI}{European Telecommunications Standards Institute}%
\acro{ER}{Effective Roughness}%
\acro{E-UTRA}{Evolved UMTS Terrestrial Radio Access}%
\acro{E-UTRAN}{Evolved UMTS Terrestrial Radio Access Network}%
\acro{EF}{Electric Field}
\acro{EMC}{Electromagnetic Compatibility}

\acro{FDD}{Frequency Division Duplexing}%
\acro{FDM}{Frequency Division Multiplexing}%
\acro{FDMA}{Frequency Division Multiple Access}%
\acro{FoM}{Figure of Merit}
\acro{FoV}{Field of View}
\acro{FSA}{Frequency Selective Absorber}
\acro{FS}{Frequency Samples}
\acro{GI}{Global Illumination} %
\acro{GIS}{Geographic Information System}%
\acro{GO}{Geometrical Optics} %
\acro{GPU}{Graphics Processing Unit}%
\acro{GPGPU}{General Purpose Graphics Processing Unit}%
\acro{GPRS}{General Packet Radio Service}%
\acro{GSM}{Global System for Mobile Communication}%
\acro{GNSS}{Global Navigation Satellite System}%
\acro{GoF}{Goodness-of-Fit}
\acro{H2D}{Human-to-Device}%
\acro{H2H}{Human-to-Human}%
\acro{HDRP}{High Definition Render Pipeline}
\acro{HSDPA}{High Speed Downlink Packet Access}
\acro{HSPA}{High Speed Packet Access}%
\acro{HSPA+}{High Speed Packet Access Evolution}%
\acro{HSUPA}{High Speed Uplink Packet Access}
\acro{HPBW}{Half-Power Beamwidth}
\acro{HA}{Horn Antenna}

\acro{IEEE}{Institute of Electrical and Electronic Engineers}%
\acro{InH}{Indoor Hotspot} %
\acro{IMT} {International Mobile Telecommunications}%
\acro{IMT-2000}{\ac{IMT} 2000}%
\acro{IMT-2020}{\ac{IMT} 2020}%
\acro{IMT-Advanced}{\ac{IMT} Advanced}%
\acro{IoT}{Internet of Things}%
\acro{IP}{Internet Protocol}%
\acro{ITU}{International Telecommunications Union}%
\acro{ITU-R}{\ac{ITU} Radiocommunications Sector}%
\acro{IS-95}{Interim Standard 95}%
\acro{IES}{Inter-Element Spacing}
\acro{IF}{Intermediate Frequency}


\acro{KPI}{Key Performance Indicator}%
\acro{K-S}{Kolmogorov-Smirnov}

\acro{LB} {Light Bounce}
\acro{LIM}{Light Intensity Model}%
\acro{LOS}{Line-Of-Sight}%
\acro{LTE}{Long Term Evolution}%
\acro{LTE-Advanced}{\ac{LTE} Advanced}%
\acro{LSCP}{Lean System Control Plane}%
\acro{LSI} {Light Source Intensity}

\acro{M2M}{Machine-to-Machine}%
\acro{MatSIM}{Multi Agent Transport Simulation}
\acro{METIS}{Mobile and wireless communications Enablers for Twenty-twenty Information Society}%
\acro{METIS-II}{Mobile and wireless communications Enablers for Twenty-twenty Information Society II}%
\acro{MIMO}{Mul\-ti\-ple-In\-put Mul\-ti\-ple-Out\-put}
\acro{mMIMO}{massive MIMO}%
\acro{mMTC}{massive Machine Type Communications}%
\acro{mmW}{millimeter-wave}%
\acro{MU-MIMO}{Multi-User MIMO}
\acro{MMF}{Max-Min Fairness}
\acro{MKED}{Multiple Knife-Edge Diffraction}
\acro{MF}{Matched Filter}
\acro{mmWave}{Millimeter Wave}

\acro{NFV}{Network Functions Virtualization}%
\acro{NLOS}{Non-Line-Of-Sight}%
\acro{NR}{New Radio}%
\acro{NRT}{Non Real Time}%
\acro{NYU}{New York University}%
\acro{N75PRP}{Near-75-degrees Partial Radiated Power}%
\acro{NHPRP}{Near-Horizon Partial Radiated Power}%

\acro{O2I}{Outdoor to Indoor}%
\acro{O2O}{Outdoor to Outdoor}%
\acro{OFDM}{Orthogonal Frequency Division Multiplexing}%
\acro{OFDMA}{Or\-tho\-go\-nal Fre\-quen\-cy Di\-vi\-sion Mul\-ti\-ple Access}
\acro{OtoI}{Outdoor to Indoor}%
\acro{OTA}{Over-The-Air}

\acro{PDF}{Probability Distribution Function}
\acro{PDP}{Power Delay Profile}
\acro{PHY}{Physical}%
\acro{PLE}{Path Loss Exponent}
\acro{PRP}{Partial Radiated Power}
\acro{PW}{Plane Wave}%
\acro{PR}{Pass Rate}%
\acro{PAS}{Power-Angle Spectrum}

\acro{QAM}{Quadrature Amplitude Modulation}%
\acro{QoS}{Quality of Service}%

\acro{RCSP}{Receive Signal Code Power}
\acro{RAN}{Radio Access Network}%
\acro{RAT}{Radio Access Technology}%

\acro{RAN}{Radio Access Network}%
\acro{RMa}{Rural Macro-cell}%
\acro{RMSE} {Root Mean Square Error}
\acro{RSCP}{Receive Signal Code Power}%
\acro{RT}{Ray Tracing}
\acro{RX}{receiver}
\acro{RMS}{Root Mean Square}
\acro{Random-LOS}{Random Line-Of-Sight}
\acro{RF}{Radio Frequency}
\acro{RC}{Reverberation Chamber}
\acro{RC-HARC}{Resonating Cavity Hybrid Anechoic-Reverberation Chamber}%
\acro{RIMP}{Rich Isotropic Multipath}
\acro{RHA}{Reference Horn Antenna}
\acro{RIMPMA}{RIMP Measurement Antenna}

\acro{SB} {Shadow Bias}
\acro{SC}{small cell}
\acro{SDN}{Software-Defined Networking}%
\acro{SGE}{Serious Game Engineering}%
\acro{SF}{Shadow Fading}%
\acro{SIMO}{Single Input Multiple Output}%
\acro{SINR}{Signal to Interference plus Noise Ratio}
\acro{SISO}{Single Input Single Output}%
\acro{SMa}{Suburban Macro-cell}%
\acro{SNR}{Signal to Noise Ratio}
\acro{SU}{Single User}%
\acro{SUMO}{Simulation of Urban Mobility}
\acro{SS} {Shadow Strength}
\acro{STD}{Standard Deviation}
\acro{SW} {Sliding Window}


\acro{TDD}{Time Division Duplexing}%
\acro{TDM}{Time Division Multiplexing}%
\acro{TD-CDMA}{Time Division Code Division Multiple Access}%
\acro{TDMA}{Time Division Multiple Access}%
\acro{TX}{transmitter}
\acro{TZ}{Test Zone}
\acro{TRP}{Total Radiated Power}


\acro{UAV}{Unmanned Aerial Vehicle}%
\acro{UE}{User Equipment}%
\acro{UI}{User Interface}
\acro{UHD}{Ultra High Definition}
\acro{UL}{Uplink}%
\acro{UMa}{Urban Macro-cell}%
\acro{UMi}{Urban Micro-cell}%
\acro{uMTC}{ultra-reliable Machine Type Communications}%
\acro{UMTS}{Universal Mobile Telecommunications System}%
\acro{UPM}{Unity Package Manager}
\acro{UTD}{Uniform Theory of Diffraction} %
\acro{UTRA}{{UMTS} Terrestrial Radio Access}%
\acro{UTRAN}{{UMTS} Terrestrial Radio Access Network}%
\acro{URLLC}{Ultra-Reliable and Low Latency Communications}%
\acro{UHRP}{Upper Hemisphere Radiated Power}%
\acro{ULA}{Uniform Linear Array}%

\acro{V2V}{Vehicle-to-Vehicle}%
\acro{V2X}{Vehicle-to-Everything}%
\acro{VP}{Visualization Platform}%
\acro{VR}{Virtual Reality}%
\acro{VNA}{Vector Network Analyzer}
\acro{VIL}{Vehicle-in-the-loop}

\acro{WCDMA}{Wideband Code Division Multiple Access}%
\acro{WINNER}{Wireless World Initiative New Radio}%
\acro{WINNER+}{Wireless World Initiative New Radio +}%
\acro{WiMAX}{Worldwide Interoperability for Microwave Access}%
\acro{WRC}{World Radiocommunication Conference}%

\acro{xMBB}{extreme Mobile Broadband}%

\acro{ZF}{Zero Forcing}

\end{acronym}

\title{A Practical Approach to Generating First-Order Rician Channel Statistics in a RC plus CATR Chamber at mmWave}
\author{A. Antón Ruiz, \IEEEmembership{Student Member, IEEE}, S. Hosseinzadegan \IEEEmembership{Member, IEEE}, J. Kvarnstrand, \IEEEmembership{Member, IEEE}, K. Arvidsson, and A. Alayón Glazunov, \IEEEmembership{Senior Member, IEEE}
\thanks{Manuscript submitted for review on February 28, 2024. The work of Alejandro Antón is supported by the European Union’s Horizon 2020 Marie Skłodowska-Curie grant agreement No. 955629. Andrés Alayón Glazunov also kindly acknowledges funding from the ELLIIT strategic research environment (https://elliit.se/).}
\thanks{A. Antón Ruiz is with the Radio Systems Group, within the Department of Electrical Engineering, University of Twente, Enschede, The Netherlands (e-mail: a.antonruiz@utwente.nl).}
\thanks{S. Hosseinzadegan is with the Electromagnetic Compatibility (EMC) team, Volvo Technology AB, Gothenburg, Sweden (e-mail: samar.hosseinzadegan@volvo.com).}
\thanks{J. Kvarnstrand is with the Method and Development Groups, Bluetest AB, Gothenburg, Sweden (e-mail: john.kvarnstrand@bluetest.se).}
\thanks{K. Arvidsson is with Bluetest AB, Gothenburg, Sweden (e-mail: klas.arvidsson@bluetest.se).}
\thanks{A. Alayón Glazunov is with the Department of Science and Technology, Linköping University, Nörrkoping Campus, Sweden (e-mail: andres.alayon.glazunov@liu.se). He is also affiliated with the Department of Electrical Engineering, University of Twente, Enschede, The Netherlands.}}
\maketitle
\begin{abstract}
This paper explores a novel hybrid configuration integrating a Reverberation Chamber (RC) with a Compact Antenna Test Range (CATR) to achieve a controllable Rician K-factor. The focus is testing directive antennas in the lower FR2 frequency bands (24.25-29.5 GHz) for 5G and beyond wireless applications. The study meticulously evaluates 39 unique configurations, using a stationary horn antenna for consistent reference K-factor characterization, and considers variables like absorbers and CATR polarization. Results demonstrate that the K-factor can be effectively adjusted within the hybrid setup, maintaining substantial margins above the noise level across all configurations. Sample independence is confirmed for at least 600 samples in all cases. The Bootstrap Anderson-Darling goodness-of-fit test verifies that the data align with Rician or Rayleigh distributions. Analysis of total received power, stirred and unstirred power and frequency-dependent modeling reveals that power variables are inversely related to frequency, while the K-factor remains frequency-independent. The hybrid RC-CATR system achieves a wide range of frequency-averaged K-factors from -9.2 dB to 40.8 dB, with an average granularity of 1.3 dB. Notably, configurations using co-polarized CATR signals yield large K-factors, reduced system losses, and improved frequency stability, underscoring the system's efficacy for millimeter-wave over-the-air testing. This research offers a cost-efficient and repeatable method for generating complex Rician fading channels at mmWave frequencies, crucial for the effective OTA testing of advanced wireless devices.

\end{abstract}


\section{Introduction}
\label{sec:introduction}
\ac{OTA} testing is now the standard for verifying the radiated communication performance of wireless devices, including smartphones, tablets, access points, base stations, and vehicle-mounted wireless equipment \cite{3GPP38827,5GAA_VATM}. It provides a realistic performance assessment in a controlled and repeatable environment. It is crucial for testing fully integrated devices and active antenna systems with multiple elements without any connectors \cite{5G_Testing_Survey}.

Wireless devices operate in complex and diverse propagation channels, but for practical and cost-effective \ac{OTA} measurements, testing should focus on a subset of relevant scenarios. The spatial characteristics of propagation channels and their interaction with antennas influence the variability in communication link quality \cite{STWC}. The \ac{PAS}, which describes the \ac{AoA} distribution of waves incident on the device, affects the fading distribution of the received signal at the \ac{DUT}. The Rician fading channel model, defined by a deterministic \ac{LOS} component and a random \ac{NLOS} component, characterizes this interaction, with the envelope of the received signal following the Rician \ac{PDF} \cite{Rician_Dist_Original}. The $K-$factor, representing the ratio of \ac{LOS} to \ac{NLOS} power, defines the model, with $K=0$ indicating a Rayleigh \ac{PDF} and $K\rightarrow\infty$ representing a deterministic, non-fading component.

\ac{mmWave} technology is crucial for \ac{5G} due to its large bandwidth availability, though it exhibits higher path loss and sparser, more directional channels compared to sub-10 GHz frequencies \cite{mmWavejust, 5Gchannels_survey}. Consequently, higher $K-$factors are more common. To overcome path loss, \ac{5G} systems employ massive \ac{MIMO} arrays with highly directive antennas, which must be tested in various channel conditions to emulate different Rician $K-$factors.

There are two limiting propagation scenarios: \ac{RIMP} and \ac{Random-LOS}. The \ac{RIMP} scenario represents an isotropic \ac{AoA} distribution, modeling the random propagation component. In contrast, the \ac{Random-LOS} scenario depicts a Dirac-delta \ac{PAS}, where deterministic waves impinge on the \ac{DUT} from specific orientations. \ac{RIMP} can be reproduced in a \ac{RC}, and \ac{Random-LOS} in anechoic or semi-anechoic chambers, given proper design \cite{LTEBOOK, RandomLOS, Kildal_hyp}. A new testing setup incorporating a \ac{CATR} system within an \ac{RC} allows for the emulation of these complex scenarios, with Rician fading with various $K-$factors \cite{BluetestCATR, EuCAP2024_ours}.

Extensive research has been conducted on characterizing the $K-$factor in \acp{RC}, aiming to achieve a low $K-$factor for accurate \ac{RIMP} emulation, which reduces uncertainty in \ac{RC} measurements like \ac{TRP} \cite{KFLit1, Kildal_RC_KF_formula, StirUnstir, ZaherVIRCKF, UncRIMP, UncRIMP2}.

Emulating Rician channels with variable $K-$factors in \acp{RC} is not new. Previous studies have explored various methods to control $K-$factors across different conditions and frequencies, typically up to 6 GHz \cite{Emul_Rician_Andres_RC, KFEmulRician, MIMORician, PprocRician, Ra_Ri_RMS, RC_Rician, ahmed2023overtheair}. However, these methods often involve post-processing or physical adjustments that are not ideal for active device testing and can lead to mechanical and repeatability challenges. Additionally, there is limited work validating that the resulting data follows a Rician distribution, a critical aspect for meaningful $K-$factor estimation.

Our approach uses a \ac{CATR} to generate a \ac{PW}, which ensures compliance with far-field requirements even for large \acp{DUT} and when high $K-$factors are desired. This setup includes a \ac{LOS} blocking plate between the \ac{RIMP} antenna and the \ac{DUT}, allowing for controlled \ac{LOS} components and lower $K-$factors in hybrid \ac{RC} and \ac{CATR} modes. This configuration ensures the \ac{DUT} experiences a consistent \ac{LOS} component within the $30$ cm ``quiet zone" of the \ac{CATR}. We also strategically place absorbers to manage power from the \ac{CATR}, enhancing repeatability and control over $K-$factor variations, especially when the \ac{DUT} is mounted on a roll tower, offering a significant improvement over previous methods \cite{RC_Rician}.

To the authors' best knowledge, this is the first time a mixed \ac{RC} plus \ac{CATR} \ac{OTA} setup is thoroughly investigated regarding the realization of a wide range of Rician $K-$factors at the \ac{mmWave} frequencies. The contributions of this paper can be summarized as follows:

\begin{itemize}
    \item We study in more depth the mixed \ac{RC} plus \ac{CATR} \ac{OTA} setup proposed in \cite{EuCAP2024_ours} as a means to produce a \ac{RC-HARC} to generate complex Rician fading channels at \ac{mmWave} frequencies by mixing the contributions of the two available channel modes, Pure-\ac{LOS} produced by the \ac{CATR} and the \ac{RIMP} produced by the \ac{RIMPMA}. This is a cost-efficient solution because the \ac{RC-HARC} is implemented as an already existing, commercially available product in a way that it was not originally designed for, without needing relevant modifications. This work analyzes not only the $K-$factor frequency response as in \cite{EuCAP2024_ours}, but also the average and frequency response of the unstirred, stirred, and total received powers, as well as the \ac{SNR}.
    \item We provide a substantial analysis of different configurations of the \ac{RC-HARC} at the lower FR2 bands for \ac{5G}, i.e., from $24.25-29.5$~GHz, achieving a wide range of frequency-averaged $K-$factors, from $-9.2$ to $40.8$~dB, with a dB-averaged granularity of $1.3$~dB, being of $5.1$~dB in the worst case. We regard the generation of environments with a wide range of $K-$factors, including large ones, as highly relevant for \ac{mmWave} \ac{OTA} testing.
    \item We consider the use of absorbers designed to attenuate the reflections coming from the \ac{CATR}, as well as attenuators and the two different polarizations of the \ac{CATR} feeder, to generate channels with different $K-$factors. All these elements are introduced in a controlled way in the setup, minimizing repeatability issues. In particular, the absorbers have a fixed mount in the chamber, and the \ac{CATR} polarization change is implemented via a switch.
    \item We propose a horn antenna as a reference for \ac{mmWave} measurements $K-$factor. We do this partly based on the recommendations of 3GPP \cite{3GPPRCStandard} guidelines in other areas, where it is specified to use a reference antenna that excites the \ac{RC} similarly to the one of the \ac{DUT}. e.g., for the spatial uniformity test. At \ac{mmWave}, antennas are expected to have directive radiation patterns and, consequently, high gains \cite{mmWavejust}. This is heavily \ac{DUT}-dependant, e.g. a base station will have a much larger gain than a mobile phone. For the proposed setup, the gain has a larger impact on the achieved $K-$factor when the \ac{CATR} excitation is used. Therefore, since directive antennas are used at \ac{mmWave} but their gain is not fixed, we decide to use a \ac{RHA} with around $14$~dBi gain \cite{DRH50} at the considered frequencies. This antenna is similar to one of the antennas used in \cite{mmWavejust} to conduct urban propagation measurements at $28$~GHz, which is also a horn antenna with $15$~dBi gain, supporting the reasonability of our choice.
    \item We follow a similar approach to what the \ac{CTIA} \cite{CTIAprechar} proposes for the assessment of uncertainties due to the lack of spatial uniformity of the \ac{RC}. \cite{CTIAprechar} requires a series of precharacterization measurements of the \ac{RC} which result in an uncertainty figure for each loading condition of the chamber. This information is stored in a table and used accordingly when taking measurements. The precharacterization measurements must be repeated only when substantial modifications are made to the chamber. In our case, we propose a precharacterization measurements assessment of the reference $K-$factor value for each of the $39$ considered configurations. The idea is to take these measurements once, store them, and select the proper configuration to obtain the desired environment, i.e., the desired $K-$factor perceived by the \ac{RHA} in the same position as it was taken during the precharacterization measurements.
    \item We check the acquired samples for independence, which is required to use the estimator from \cite{KFEmulRician}, as well as for the proper application of the \ac{GoF} tests.
    \item We check that the acquired data follows a Rician distribution, making the $K-$factor evaluation meaningful. We use a bootstrap-based \ac{AD} \ac{GoF} test, which is the best way found in the literature to check for \ac{GoF} for Rician distribution without prior knowledge of the distribution's parameters. In addition, it is used to check for \ac{GoF} for Rayleigh distribution. To the authors' best knowledge, this is the first time it is applied to statistical data obtained from \ac{RC} measurements.
\end{itemize}

The remainder of the paper is organized as follows: Section~\ref{section: Setup} presents the setup for the experiments, presenting the \ac{RC-HARC} chamber, the \ac{RHA}, the considered configurations of the chamber, as well as the connections of the different cables and ports and the \ac{VNA} configuration. Section~\ref{section: Methodology} presents the methodology used for the analysis of the measured data, including the assumed Rician distributed signal model, the method used to check for the independence of the samples, the procedure to obtain the estimate the received power from $S_{21}$ data, the \ac{SNR} estimation method, the chosen $K-$factor estimator, and the \ac{AD} bootstrap-based \ac{GoF} test used. Section~\ref{section: Results} presents the results based on the methodology described in Section~\ref{section: Methodology}, therefore presenting the independence of samples, average \ac{SNR} and power, $K-$factor, including its frequency dependence, and the results of the \ac{GoF} tests. Finally, Section~\ref{section: Conclusions} presents the conclusions after analyzing the results from Section~\ref{section: Results}, as well as the limitations and future work.

\section{Experimental setup and Measurement cases}
\label{section: Setup}
This section describes the measurement setup to generate the \ac{RIMP}, the Pure-\ac{LOS}, and the compound propagation channel.
\subsection{\ac{RC-HARC} Chamber}

\begin{figure*}[!t]
\centering
\includegraphics[width=2\columnwidth]{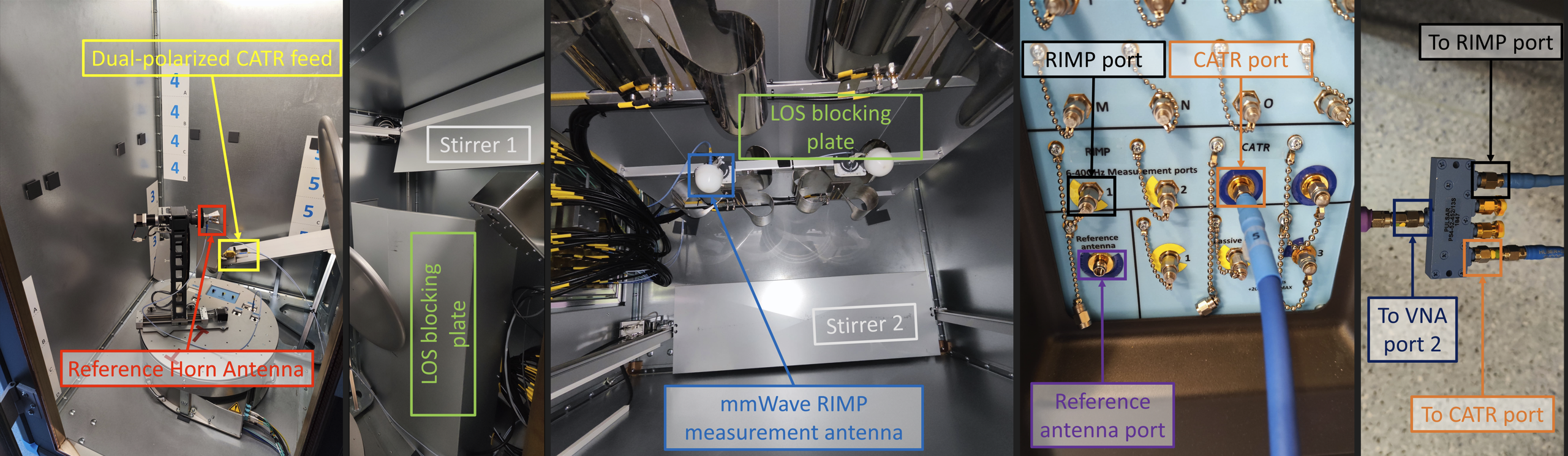}
\caption{From left to right, Bluetest RTS65 with the \ac{CATR} option and without any absorbers installed, corresponding to the ``NoAs" case, depicting the \ac{RHA} and the dual-polarized \ac{CATR} feed. Then we have the Stirrer 1 and the \ac{LOS} blocking plate, which hides the \ac{mmWave} \ac{RIMPMA}, depicted in the next picture, which is a ceiling shot taken from behind the \ac{LOS} blocking plate, where the Stirrer 2 can be seen. Then, we have the front panel of the \ac{RC}, with the three used ports highlighted: \ac{RIMP} port, \ac{CATR} port, and reference antenna port. Finally, the 1:4 splitter goes outside the chamber and indicates where each port is connected. Note the terminations in the unused ports. Note also that the splitter configuration shown is for the RaC case. In R and C cases the port of the splitter going to the \ac{CATR} or \ac{RIMP} ports, respectively, would be terminated too.}
\label{F1}
\end{figure*}
\begin{figure}[!t]
\centering
\includegraphics[width=0.8\columnwidth]{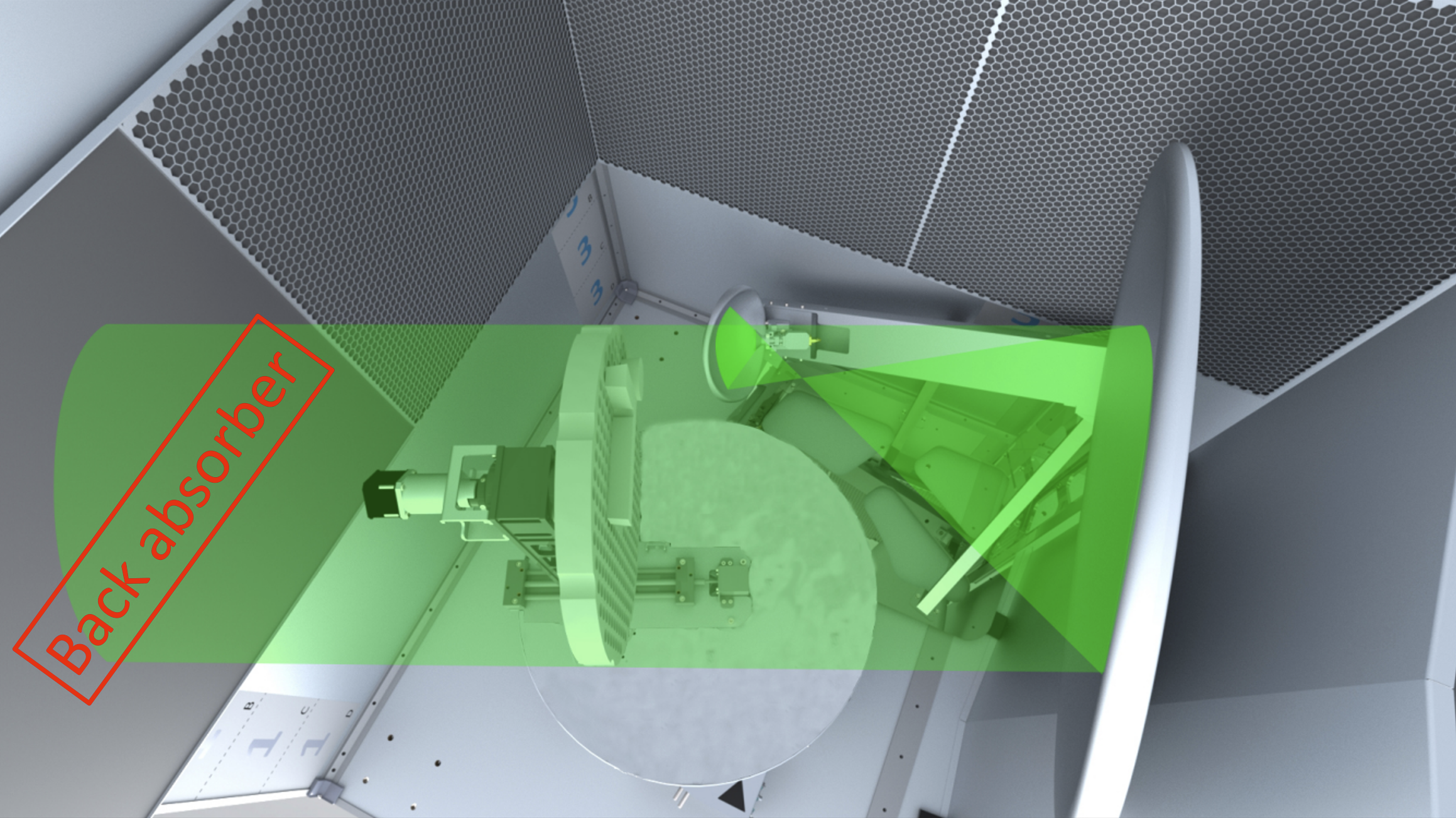}
\caption{CAD render showing geometry and signal flow (in green). The ``quiet zone" is located within the lightest green volume. Notice how all absorbers are installed in the chamber, corresponding to the ``AAs" case, including the back absorber, depicted in red, which intercepts the signal flow.} Source: \cite{BluetestCATR}.
\label{F2}
\end{figure}
\begin{figure}[!t]
\centering
\includegraphics[width=0.8\columnwidth]{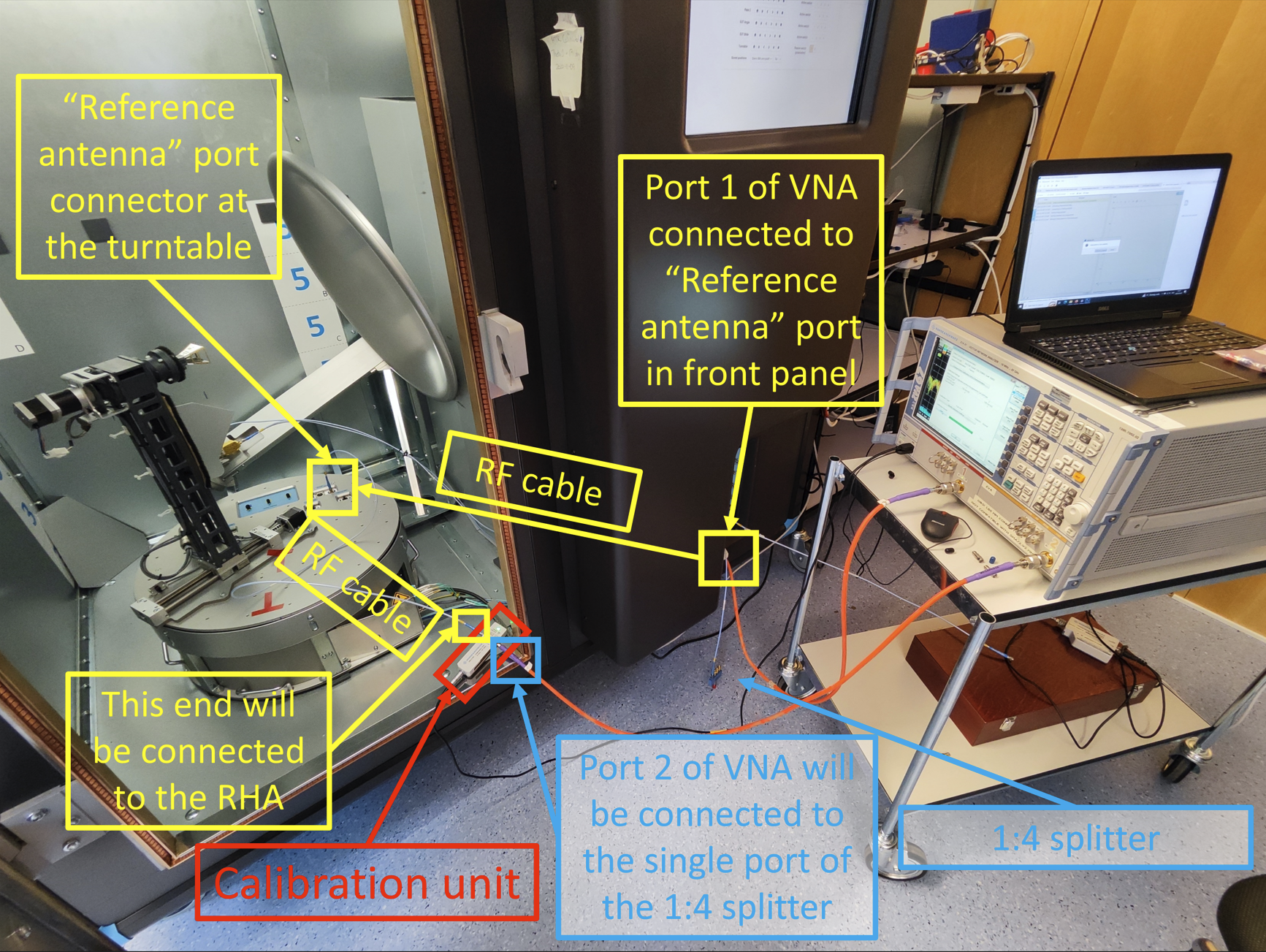}
\caption{Experiment setup for calibrating the \ac{VNA}.}
\label{F3}
\end{figure}

Fig.~\ref{F1} shows the experimental setup showing the interior of a Bluetest RTS65 \ac{RC} chamber with the \ac{CATR} option installed, with dimensions of the chamber of $1945\times2000\times1440$~mm$^3$ (WxHxD) \cite{BluetestCATR}. The chamber is designed to be used as a compact 2-in-1 test system, providing either a \ac{RIMP} or a Pure-\ac{LOS} only field in the test zone when it is excited through the \ac{RIMP} or the \ac{CATR} port, respectively. Hence, \ac{TRP} and radiation pattern measurements can be performed within the same space since the \ac{CATR} mode uses absorbers on the walls. For the \ac{RIMP} operation, the chamber has a total of 3 physical stirring mechanisms. One of them is the turntable, which is not used in this work. The other two stirrers, which are linear, are shown in Fig.~\ref{F1} as Stirrer 1 and Stirrer 2. They operate as just one stirring mechanism, moving coordinately.

The \ac{CATR} generates a \ac{PW} with a dual parabolic reflector illuminated by a dual-polarized antenna feed shown in yellow in Fig.~\ref{F1}, and it has a $0.6$~dB amplitude and a $4^\circ$ phase ripples. The performance is maintained from $24.25-42$~GHz within a $30$~cm diameter cylindrical ``quiet zone" shown in Fig.~\ref{F2} \cite{CATR}. It must be noted that the alignment between the \ac{DUT} and the \ac{CATR} will have a relevant effect on the realized $K-$factor, while the position of the \ac{DUT}, as long as it is within the ``quiet zone", will have a smaller impact since the direct coupling between the \ac{CATR} and a \ac{DUT} pointing at it has a small variation within the ``quiet zone". The polarization of the feeder antenna of the \ac{CATR} can be selected using a passive switch or connecting to the other \ac{CATR} port. 

The reflections coming from the reflector or the \ac{DUT} are reduced by carbon-loaded foam absorbers placed around the chamber, as shown in Fig.~\ref{F2}. Except for the absorber on the wall towards which the \ac{PW} is directed (``back absorber"), all the other absorbers are \ac{FSA}. They are covered by a metallic honeycomb pattern with a size of the periodic metallic pattern such that it is small compared to the wavelength at sub-$6$~GHz frequencies for which the chamber is also designed to operate in \ac{RIMP}, thus providing a strong reflection. However, at \ac{mmWave}, the periodic metallic pattern size is larger than the wavelength. Therefore providing a small reflection and letting the signals reach the carbon-loaded absorber behind the honeycomb. The back absorber has the same material and size as the \ac{FSA}, except that the metallic honeycomb part is removed, thus providing less reflection. This is because most of the power coming from the reflector is directed to this back absorber.

\subsection{Considered Measurement Configurations}
A total of 39 measurement cases were considered, by combining 4 configuration variables in the setup specified below.

\begin{itemize}[label=$\blacksquare$]
    \item \ac{RC}, \ac{CATR} or mixed mode of the \ac{RC-HARC} chamber are used. As depicted in Fig.1, we use the \ac{RIMP} and \ac{CATR} ports of the chamber to excite either the \ac{RIMPMA} or the dual-polarized \ac{CATR} feed, respectively. The first will emulate the \ac{RIMP} environment, generating multiple \ac{PW}s. The second will emulate the Pure-\ac{LOS} environment (a single \ac{PW}), only in case of installing all the absorbers. If not, it will generate an environment with a direct \ac{PW} and then some reflections that might or might not interact with the stirrers. Both ports can be used simultaneously using a splitter as shown in Fig.~\ref{F1}. Therefore, the possible configurations are:
    \begin{itemize}
        \item R - only \ac{RIMP} port connected to the splitter to generate the \ac{RIMP} channel.
        \item C - only \ac{CATR} port connected to the splitter to emulate a Pure-\ac{LOS} channel when all absorbers are placed or a Rician channel.
        \item RaC - \ac{RIMP} and \ac{CATR} ports connected to the splitter. A mixture of stirred and unstirred components produces a Rician channel with desired $K-$factors.
    \end{itemize}
    \item \ac{CATR} dual-polarized feeder. The feeder has two orthogonal polarizations that can be selected using a passive switch that routes the signal to each feeder port. A single polarization is used at a time in our work.
    \begin{itemize}
        \item PSX - is stated in \ac{RIMP} only cases in which the position of the passive switch is irrelevant since the \ac{CATR} port is not used.
        \item PS1 - passive switch in position 1 corresponding to the co-polarized field component of the \ac{RHA} in the experiment's setup. This produces a strong \ac{LOS} coupling of the \ac{CATR} signal to the \ac{RHA}.
        \item PS2 - passive switch in position 2 corresponding to the cross-polarized to the \ac{RHA} in the experiment's setup. This produces a weak \ac{LOS} coupling of the \ac{CATR} signal to the \ac{RHA}.
    \end{itemize}
    \item Absorbers are used in three configurations.
    \begin{itemize}
        \item NoAs - no absorbers with minimal attenuation of both \ac{RIMP} and \ac{CATR} signals.
        \item BAs - back absorber only, as shown in Fig.~\ref{F1}, with a large reduction of the stirred component coming from the reflections of the \ac{CATR} signal.
        \item AAs - all absorbers present, as shown in Fig.~\ref{F3}, with a large reduction of the stirred component in general.
    \end{itemize}
    \item Attenuators are used in the cases where \ac{RIMP} and \ac{CATR} ports are used at the same time (i.e., the ``RaC" cases). The attenuators allow changing the realized $K$-factor, with minimal changes to the setup. The adapters ($2.92$~mm male to $2.4$~mm female and $2.4$~mm male to $2.92$~mm female) are required for the ``X0ATR" configurations but not for the ``X0ATC" cases due to the used components and arrangement of ports of the chamber. More on this can be found in Section~\ref{section: Cabling}.
    \begin{itemize}
        \item NAT - no attenuators.
        \item 10ATC - 10 dB attenuator on \ac{CATR} branch.
        \item 20ATC - 20 dB attenuator on \ac{CATR} branch.
        \item 10ATR - 10 dB attenuator plus adapters on \ac{RIMP} branch.
        \item 20ATR - 20 dB attenuator plus adapters on \ac{RIMP} branch.
    \end{itemize}
\end{itemize}

\begin{table*}
\centering
\caption{Relevant statistics of $K-$factor, stirred, unstirred and total powers.}
\label{T1}
\includegraphics[width=2\columnwidth]{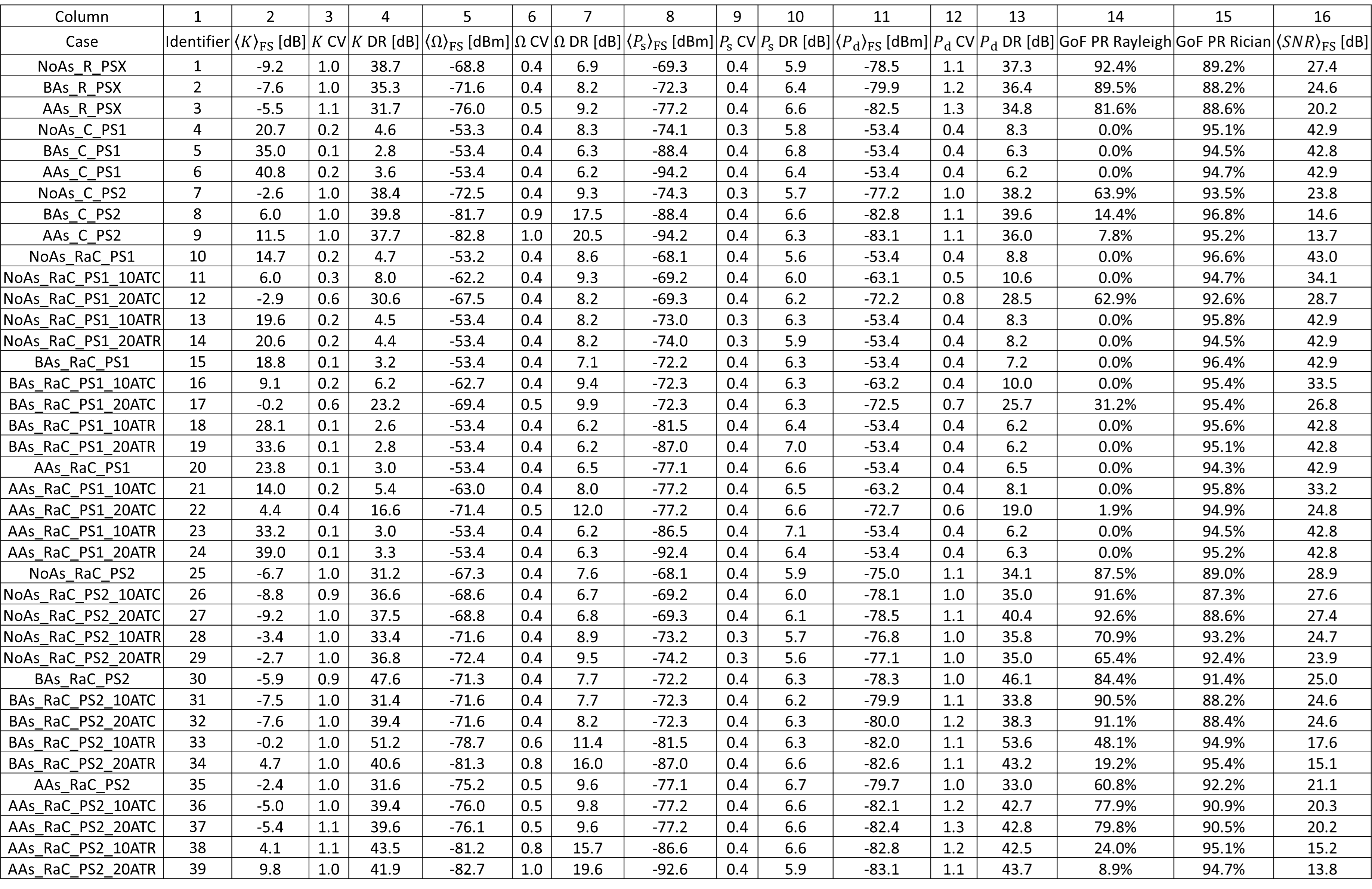}
\end{table*}

An exhaustive list of the cases can be found in Table~\ref{T1}, where each case has been assigned to a numeric identifier, which will be used in the figures to ease interpretation.

\subsection{Instruments and Measurements}

\subsubsection{Cabling}
\label{section: Cabling}
To perform this experiment, a $2.92$~mm 1:4 splitter was used. The \ac{VNA} has $1.85$~mm ports, to which $2.4$~mm cables are connected directly. The cable connected to port 2 of the \ac{VNA} is connected to the single port of the splitter with a $2.4$~mm female to $2.92$~mm male adapter.
For the ``R" cases, three of the ports of the splitter are terminated, and the other is connected via a $2.92$~mm cable to the ``RIMP" port of the chamber ($2.92$~mm) that reaches the \ac{RIMPMA}. For the ``C" cases, three of the splitter ports are terminated. The other port is connected via a $2.92$~mm male to $2.4$~mm female adapter and a $2.4$~mm cable to the ``CATR" port of the chamber ($1.85$~mm), that reaches the dual-polarized \ac{CATR} feed. 
The ``RC" cases are a mixture of ``R" and ``C", where the splitter is set up as depicted in Fig.~\ref{F1}. Then the port 1 of the \ac{VNA} is connected to the ``Reference antenna" port of the chamber, that reaches the ``Reference antenna" port at the turntable, which is then connected to the \ac{RHA} via a $2.4$~mm cable. 

\subsubsection{VNA and Measurement Configuration}
\label{section: VNA}
The \ac{VNA} was configured to perform an $S_{21}$ frequency sweep from $24.25-29.5$~GHz with $10$~MHz steps according to the FR2 bands resulting in $526$ \ac{FS} for each of the $39$ cases. An \ac{IF} bandwidth of $1$~KHz was used, which provides a noise level below $-82$~dBm according to the manufacturer. The output power of the \ac{VNA} was set to $10$~dBm. The \ac{VNA} calibration plane was, on the one side, at the port of the \ac{RHA} and, on the other side, at the $2.4$~mm female port of the adapter connected to the single port of the 1:4 splitter. This is depicted in more detail in Fig.~\ref{F3}.

For each \ac{FS}, $600$ samples were collected. The samples were taken at $600$ unique positions of the two mode-stirrers of the chamber (see Fig.~\ref{F1}) while keeping the turntable fixed during all measurements to ensure that the \ac{RHA} was pointed towards the reflector. Each of the $600$ positions of the stirrers is the same for all frequencies since the movement of the stirrers is done in the outer loop, being the frequency sweep in the inner loop.

\subsection{Reference Antenna}

All measurements have been done with a linearly-polarized double ridged horn antenna, designed to operate from $4.5-50$~GHz, whose specs can be found in \cite{DRH50}. The \ac{RHA} is always placed statically (see Fig.~\ref{F1}), such that the turntable will not rotate. Hence, the turntable stirring is off. The chosen position aligns the max gain of the \ac{RHA} and its polarization with the \ac{CATR}; hence both the co- or cross-polarization can be measured. This serves our purposes well because a more controlled interaction between the \ac{RHA} and the \ac{CATR} can be achieved, thus achieving large $K-$factors. The horn antenna has been chosen for reference measurements because it's a directional antenna with a well-known pattern and accessible \ac{RF} ports, allowing complex-valued $S_{21}$ measurements. $K-$factor can then be straightforwardly estimated as discussed further in Section~\ref{section: Rician}. 

\section{Analysis Methodology}
\label{section: Methodology}
This section presents the assumed Rician fading \ac{PDF} and the corresponding parameters to emulate the first-order statistics in the \ac{RC-HARC}. We present the method to determine the number of independent samples and estimate the \ac{SNR}, the average received power $\Omega$, and the $K-$factor. The \ac{GoF} test used to determine whether the generated signals belong to the Rician \ac{PDF} is also presented.
\subsection{Rician Signal Model}
\label{section: Rician}
The Rician distribution of the envelope of the complex signal amplitude received by the antenna $|v|$ is given by \cite{Rician_Dist_Original},
\begin{eqnarray}\label{eq:5x}
f_{|v|}(|v|)&=&\frac{2\left(1+K\right)|v|}{\Omega}\exp\left(-K-\frac{(K+1)|v|^{2}}{\Omega}\right)\nonumber \times \\ &&I_{0}\left(2\sqrt{\frac{K(1+K)}{\Omega}}|v|\right),
\end{eqnarray}
where $v=S_{21}$ is, in our case, the measured quantity in the chamber, and $I_{0}$ is the modified Bessel function of the first kind and zeroth order. The \ac{PDF} parameters are the Rician $K-$factor defined as
\begin{equation}
K=\frac{P_\mathrm{d}}{P_\mathrm{s}},\label{eq:1x}
\end{equation}
where $P_\mathrm{d}$ and $P_\mathrm{s}$ are the powers of the \ac{LOS} and the \ac{RIMP} components, respectively. More specifically, we are considering $P_\mathrm{d}$ as all the power that does not interact with the mode stirrers of the chamber. On the other hand, $P_\mathrm{s}$ is the power that interacts with the chamber's stirrers, and we will assume that results from an isotropic wave field distribution, thus generating a \ac{RIMP} field. The $K-$factor measures the severity of field fluctuations generated for a specific scenario. The second parameter is the total received power
\begin{equation}
\Omega=P_\mathrm{d}+P_\mathrm{s},\label{eq:2x}
\end{equation}
i.e., the average received power. The two limiting cases, the \ac{RIMP} and the Pure-\ac{LOS} channels, directly arise from $K\rightarrow 0$ and $K\rightarrow\infty$, respectively. The former turns into the Rayleigh \ac{PDF} \cite{RC_Rician,LTEBOOK}, while the latter case turns into the Dirac-$\delta$ function \cite{Rician_Dist_Original}. Intermediate values of $K$ will describe Rician propagation channels between the RIMP and the LOS channels. 


\subsection{Independence of the samples}
\label{SIndep}

The independence of the samples is relevant for several aspects of this work, including the \ac{GoF} tests and the $K-$factor estimation. Following the discussion in \cite{CorrMatrix, CorrMat2}, the number of independent or effective samples $N_\mathrm{eff}$ in the presence of more than one mode stirrer should be computed through the use of a circular-shift correlation matrix. The threshold used in \cite{CorrMatrix, CorrMat2} for considering the correlation between samples is consistent with the one present in the standards \cite{IEC61000421,3GPPRCStandard}. If fully satisfied, it ensures, with a $95\%$ likelihood and for sample sizes larger than 100, that all samples are independent. In the case of this work, there are two mode stirrers, although they move at the same time at equispaced steps. $N_\mathrm{eff}$ is evaluated according to (3) from \cite{CorrMatrix}, where $N=600$ are samples corresponding to unique positions of the mode stirrers and measured for each of the $526$ \ac{FS}. 

\subsection{Received Power Estimation}

As discussed in Section~\ref{section: VNA}, the \ac{VNA} was configured to perform a frequency sweep reporting the calibrated complex-valued $S_{21}$. To assess the \ac{SNR}, it is necessary to know the measured power of the receiving port of the \ac{VNA} for each sample and the actual noise level of the instrument at the receiving port.

To determine the actual measured power at the \ac{VNA} receiving port, we accounted for all compensated losses and measured them along with the output power of the \ac{VNA} port. Consequently, the received power at the \ac{VNA} port is calculated as follows.
\begin{equation}
P_\mathrm{rec}[\mathrm{dBm}]=P_\mathrm{out}[\mathrm{dBm}]-L_\mathrm{cal}[\mathrm{dB}]-L_\mathrm{eff}[\mathrm{dB}]+S_{21}[\mathrm{dB}],
\label{Eq2}
\end{equation}
where the notation used indicates that the variable $X[\mathrm{dB}]$ denotes the dB value of the variable $X$ in linear units. Furthermore, the meaning of the terms in \eqref{Eq2} is explained below.
\begin{itemize}
    \item $L_\mathrm{cal}[\mathrm{dB}]$ denotes the total losses of all the cables and connectors accounted for in the calibration (see Fig.~\ref{F3}).
    \item $L_\mathrm{eff}[\mathrm{dB}]$ denotes the losses of the reference antenna due to its efficiency, which are compensated by measurement software.
    \item $P_\mathrm{out}[\mathrm{dBm}]$ is set to $10$~dBm, which, according to the calibration report of the instrument, can be output for the considered frequency range. Therefore, we use $10$~dBm across all the frequency ranges.
    \item $S_{21}[\mathrm{dB}]=20\log_{10}{\left|S_{21}\right|}$ is the link (receive) power between the transmit and receive port in the absence of the rest of considered losses and \ac{VNA} output power.
\end{itemize}

Hence, an estimate of the total (average) received power at a single frequency is estimated by 
\begin{equation}
\hat{\Omega}=\langle P_\mathrm{rec} \rangle_{\mathrm{SP}},
\end{equation}
where $P_\mathrm{rec}$ is the received power \eqref{Eq2} in linear units [$\mathrm{mW}$], and the brackets $\langle \bullet \rangle_{\mathrm{SP}}$ denote sample averaging over the $N_\mathrm{eff}$ stirrer positions (SP).

\subsection{SNR Estimation}
The noise level of the \ac{VNA} $NL[\mathrm{dBm}]$ was estimated for each $526$ \ac{FS}. The measurement was conducted by terminating both \ac{VNA} ports with $50$~$\Omega$ loads and then measuring the $S_{21}$ with the same \ac{IF} bandwidth of $1$~kHz and an output power of $0$~dBm. Hence, with calibration turned off, the measured $|S_{21}|^2$ in dB is equivalent to a noise realization's power level in dBm. For each of the $526$ \ac{FS}, $1000$ samples of $S_{21}$ (noise realizations) were collected, with a sampling time spacing of $10$~ms. Then, the noise level at a given frequency was determined as the average of those $1000$ noise samples (NS), i.e., $\langle |S_{21}|^2 \rangle_{\mathrm{NS}}$. Each \ac{FS} single noise level value is denoted as $NL[\mathrm{dBm}]=10\log(\langle |S_{21}|^2 \rangle_{\mathrm{NS}})$. The average noise level of the \ac{VNA} over the considered frequency range from $24.25-29.5$~GHz with this configuration is $\langle NL \rangle_{\mathrm{FS}}[\mathrm{dBm}]=-96.3$~dBm, which is below the $-82$~dBm that the instrument manufacturer guarantees. The peak-to-peak variation of $NL[\mathrm{dBm}]$ is $4.1$~dB.

With the average received power and the noise level, the \ac{SNR} at stirrer position $i_\mathrm{SP} \in \left[1,600\right]$ and \ac{FS} $i_\mathrm{FS} \in \left[1,526\right]$ can be computed as 
\begin{equation}
SNR\left(i_\mathrm{SP},i_\mathrm{FS}\right)[\mathrm{dB}]=P_\mathrm{rec}\left(i_\mathrm{SP},i_\mathrm{FS}\right)[\mathrm{dBm}]-NL\left(i_\mathrm{FS}\right)[\mathrm{dBm}].
\label{Eq6}
\end{equation}
It must be noted that \eqref{Eq6} can be less than $0$ if two conditions are met: first, if $S_{21}[\mathrm{dB}]$ is as low that $P_\mathrm{rec}[\mathrm{dBm}]$ is dominated by the noise level, i.e., if $S_{21}[\mathrm{dB}] \rightarrow -\infty$) and, second, if the noise realization for that particular SP sample $i_\mathrm{SP}$ and at a particular \ac{FS} $i_\mathrm{FS}$ is below the average noise level at that \ac{FS}. 
Therefore, looking at the average \ac{SNR} rather than the sample by sample \ac{SNR} is more relevant. Hence $SNR_\mathrm{avg}[\mathrm{dB}]$ is defined as the average \ac{SNR} in dB of a given configuration
\begin{equation}
SNR_\mathrm{avg}[\mathrm{dB}]=\langle\langle P_\mathrm{rec}\rangle_\mathrm{SP} \rangle_\mathrm{FS} [\mathrm{dBm}]-\langle NL \rangle_\mathrm{FS}[\mathrm{dBm}],
\label{Eq5}
\end{equation}
where the operators $\langle \bullet \rangle_{\mathrm{SP}}$ and $\langle \bullet \rangle_{\mathrm{FS}}$ denote sample averaging over the stirrer positions (SP) and frequency samples (FS), respectively. The averaging is applied to the linear values, not the ones in dB.

\subsection{$K-$factor Estimation}
\begin{figure}[!t]
\centering
\includegraphics[width=1\columnwidth]{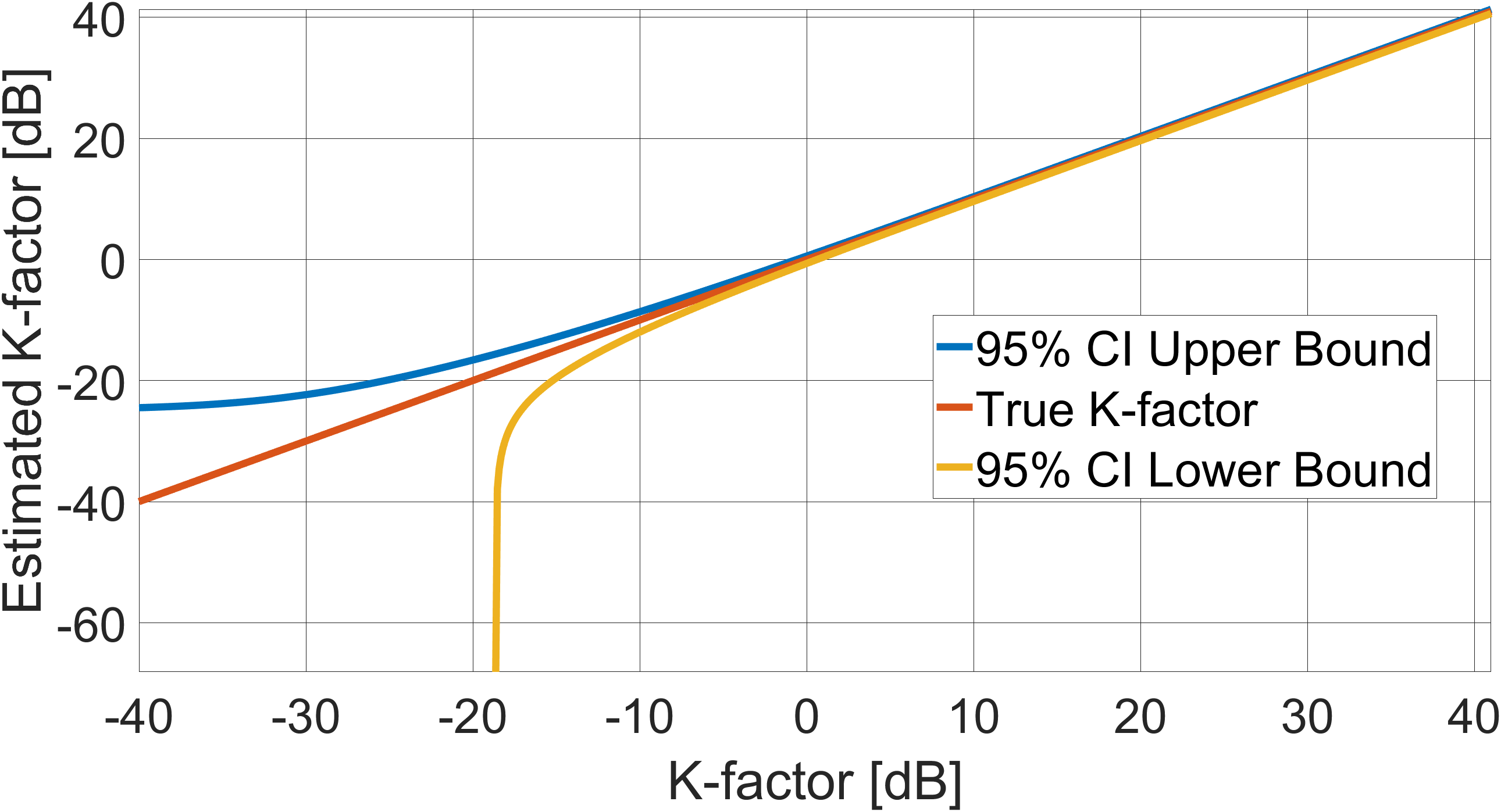}
\caption{$95\%$ \ac{CI} for the $K$-factor estimator used in this work (from \cite{KFEmulRician}) and $600$ samples or stirrers' positions.}
\label{F4}
\end{figure}
We use the $K-$factor estimator presented in \cite{KFEmulRician} because it is unbiased, which, for the sake of completeness, we reproduce here
\begin{eqnarray}
     \hat{K}&=&\frac{N_\mathrm{eff}-2}{N_\mathrm{eff}-1}\hat{K_2} -\frac{1}{N_\mathrm{eff}}, \label{Eq8a}\\
     \hat{K_2}&=&\frac{|\langle \tilde{S}_{21} \rangle_{\mathrm{SP}}|^2}{\langle|\tilde{S}_{21}|^2\rangle_{\mathrm{SP}}-|\langle \tilde{S}_{21}\rangle_{\mathrm{SP}}|^2},\label{Eq8b}
\end{eqnarray}
where 
\begin{eqnarray}
     \tilde{S}_{21}&=&{S}_{21}\sqrt{\frac{\hat{\Omega}}{\langle|S_{21}|^2\rangle_{\mathrm{SP}}}}, \label{Eq8c}
\end{eqnarray}
is the normalized transfer function, $N_\mathrm{eff}$ is the number of independent samples, which will be proven to be equal to the number of stirrers' positions ($600$) in Section~\ref{section: indep}. In \cite{KFestim_negatives}, it was observed that due to its unbiased nature, it could lead to negative (in linear units) estimates of low $K-$factors when the number of samples is limited. In our case, we have $N_\mathrm{eff}=600$, which makes the $95\%$ \ac{CI} hit an asymptote on its lower bound when the actual $K$-factor is lower than, approximately, $-18$~dB, as can be observed in Fig.~\ref{F4}. In this work, for the \ac{FS} of each case where the estimated $K$-factor is negative (in linear units), we disregard that \ac{FS} of that case from all the analysis of the considered parameters ($\Omega$,$K$,$P_{\mathrm{s}}$ and $P_{\mathrm{d}}$). The percentage of \ac{FS} where the estimated $K-$factor is negative is $0.3\%$ on average across all cases and, for an individual case, the highest percentage of \ac{FS} where the estimated $K-$factor is negative is $1.3\%$. The expressions to obtain the $95\%$ \ac{CI} are (43) and (44) from \cite{KFEmulRician}.

\subsection{Goodness-of-Fit Testing}
\ac{GoF} tests assess whether a given dataset comes from a hypothesized distribution. This work considers two separate null hypotheses: whether the data follows Rayleigh \ac{PDF}, while the second is whether the data follows the Rician \ac{PDF}. This type of \ac{GoF} problem is a composite one or, equivalently, $H_0$ is composite. The alternative hypothesis $H_1$ is that the distribution followed by the data belongs to a different family than the one specified by $H_0$. For the so-called type I error, i.e., the likelihood of incorrectly rejecting $H_0$, $\alpha=0.05$ is set, meaning that the \ac{GoF} test has a $5\%$ likelihood of rejecting $H_0$ when it is true. On the other hand, for the so-called type II error, i.e., failing to reject $H_0$ when it is false, the likelihood of the event is $\beta$. The power of the test is defined as $1-\beta$. Ideally, $\alpha$ and $\beta$ shall be as close to 0 as possible. However, that is not possible due to the finite nature of the number of independent samples. Indeed, $\alpha$ and $\beta$ are inversely related, so, for example, decreasing $\alpha$ would increase $\beta$, thus decreasing the power of the test, and vice versa. The power of the test can only be computed when $H_1$ is specified \cite{Chisq}, which is not the case in this work, and testing it for an array of relevant distributions falls beyond the scope of the work.

The data obtained here is continuous, so the chi-square \ac{GoF} would be disregarded. However, according to \cite{Chisq}, a modified version of the chi-square test can be applied to continuous data with generally good performance. Nevertheless, the performance of the \ac{AD} test is generally better than the chi-square test, according to their results. In addition, the \ac{AD} \ac{GoF} test has been successfully used for \ac{RC} applications \cite{ADGoF}, and therefore we use it.

Now moving to the composite $H_0$, it must be noted that, if one recklessly applies a \ac{GoF} test applicable to a simple $H_0$ in the case of a composite $H_0$, it will lead to an increased probability of accepting $H_0$ \cite{ADGoF}, or, similarly, a decrease of the power of the test. This would make the test useless since it will not reject $H_0$ when it is false as much as one would expect from the power of the test for simple $H_0$. To solve this, there are tables of modified critical values for some \ac{GoF} tests. However, while there are tables of critical values for the composite \ac{GoF} test when $H_0$ is that the distribution is an exponential \cite{ADStephens} (or Rayleigh, for which the same values apply \cite{ADGoF}), there are not, to the best of the authors' knowledge, for the case of the Rician distribution. For these cases, resorting to bootstrap-based \ac{GoF} tests is necessary \cite{BootstrapGoF}.

Hence, a bootstrap-based \ac{AD} \ac{GoF} test with $\alpha$ set to $0.05$ is used in this work. $H_0$ is set to be either a Rayleigh or a Rician distribution, i.e., two different \ac{GoF} tests are applied for every of the $526$ \ac{FS} that contain $600$ samples for each of the $39$ considered configurations. The data used as input for each of the \ac{GoF} tests is composed of the $600$ samples of a given \ac{FS} and a given configuration (e.g., ``BAs\_RaC\_PS1"). The implementation of the bootstrap method is based on a modification of the ``adtest" function from MATLAB to use it for Rayleigh and Rician distributions with composite $H_0$. The original ``adtest" function lacks the possibility of being used as a \ac{GoF} test for Rayleigh and Rician distributions, so we have introduced the necessary changes and new code to make it work for such distributions. This function contains the function ``adtestMC", which we also modify and it simulates the critical values and p-values for the \ac{AD} test using a Monte Carlo simulation. Instead of setting a fixed number of bootstrap samples (``B" from \cite{BootstrapGoF}), it sets them to make the standard error of the estimated p-value lower than the parameter ``mctol", which we set to $0.01$, being this the default value of ``mctol", which is present in the original ``adtestMC" function.

\section{Results and analysis}
\label{section: Results}
In this section, we present the results of the parameters specified in Section~\ref{section: Methodology}.
\subsection{Independence of the samples}
\label{section: indep}
Applying the methodology from Section~\ref{SIndep}, we found that all the collected samples can be considered independent. Namely, $N_\mathrm{eff}=N=600$, i.e., all collected samples are independent for all considered chamber configurations and frequencies.

\subsection{\ac{SNR}}

The frequency-averaged \ac{SNR} or $\langle SNR\rangle_\mathrm{FS}$ is at least $13.7$~dB and up to $43$~dB, as shown in column $16$ of Table~\ref{T1}. This is a good margin to the average noise floor; therefore, this, together with the sample independence, leads us to assume that all the measured samples can be used in our analysis. This does not mean that they are noise-free. Indeed, if we consider the \ac{SNR} at a particular frequency, there are some cases in which the average $\langle\ac{SNR}\rangle_\mathrm{SP}$ is rather low. For example, as low as $1.11$~dB for a \ac{FS} of case $9$. However, this a limited occurrence, since, e.g., the percentage of frequencies at which the \ac{SNR} is $5$~dB or more, is only less than $100\%$ for cases $9$ ($89\%$) and $39$ ($95\%$). Moreover, the percentage of frequencies at which the \ac{SNR} is $10$~dB or more, is only less than $100\%$ for cases $8$ ($83\%$), $9$ ($65\%$), $34$ ($93\%$), $38$ ($94\%$) and $39$ ($66\%$).

\begin{figure*}
\centering
\subfloat[$\Omega$][\label{F5a}]{\includegraphics[width=1\columnwidth]{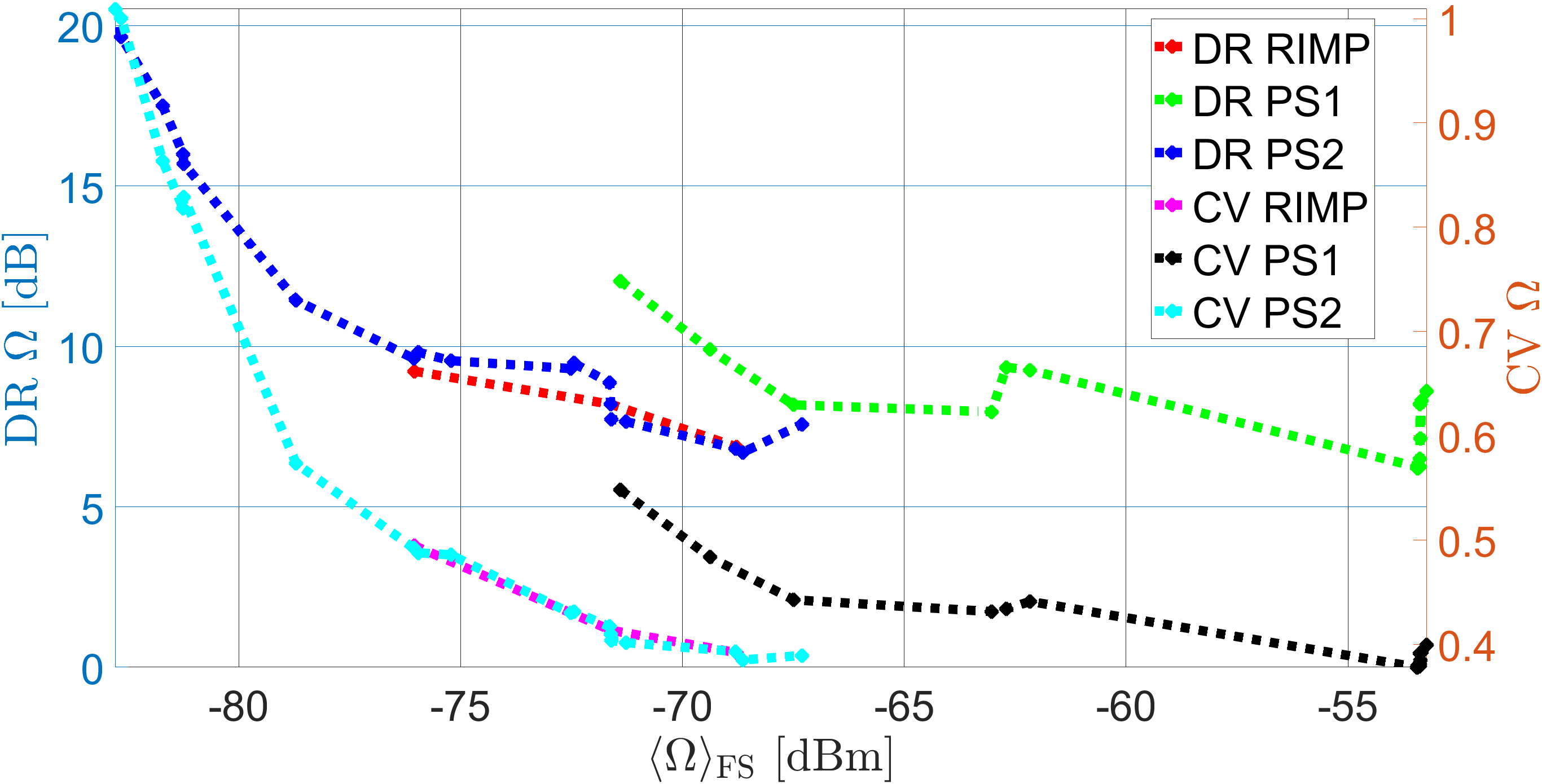}}
\hfil
\subfloat[$K$][\label{F5b}]{\includegraphics[width=1\columnwidth]{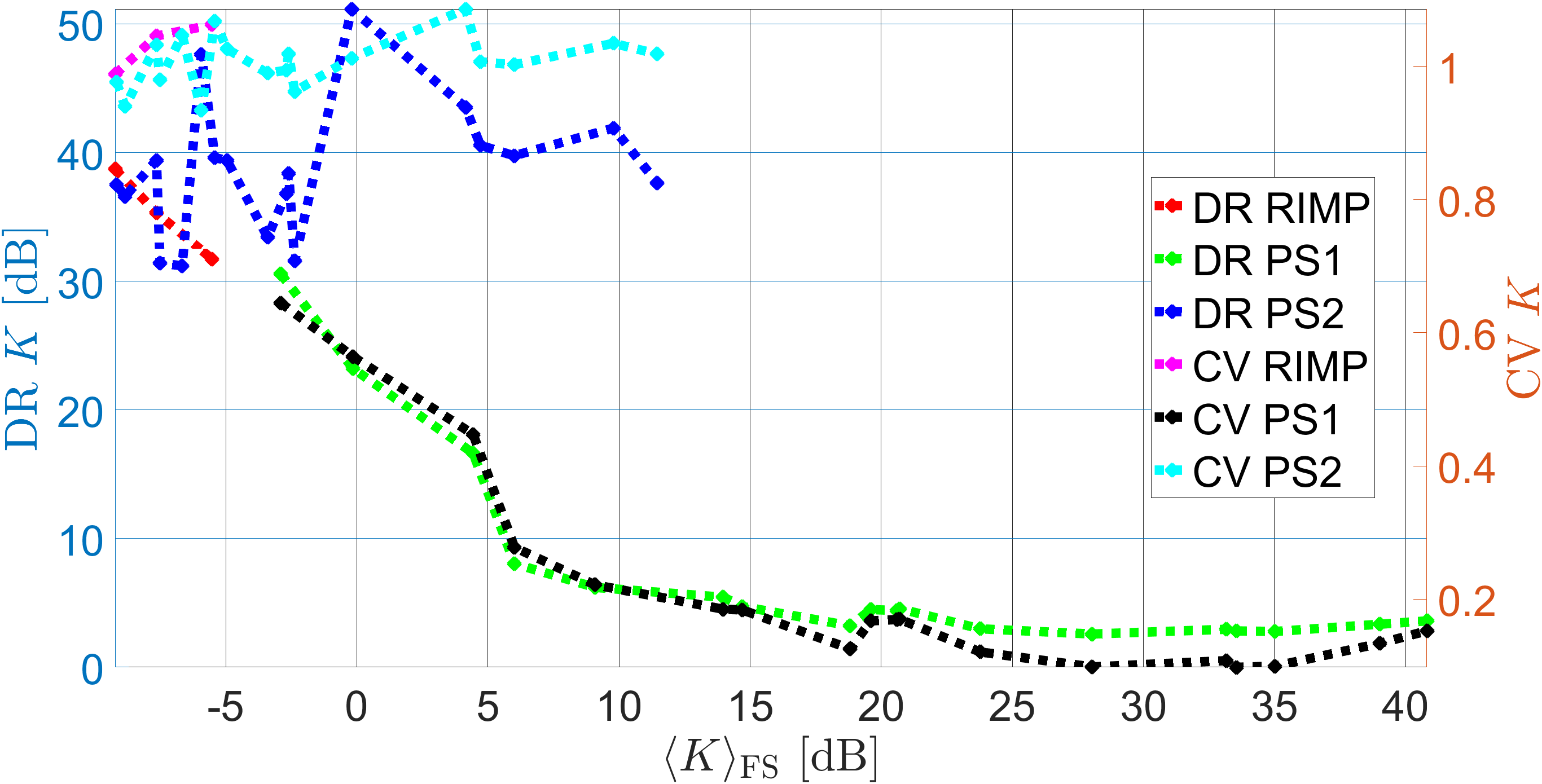}}
\hfil
\subfloat[$P_\mathrm{s}$][\label{F5c}]{\includegraphics[width=1\columnwidth]{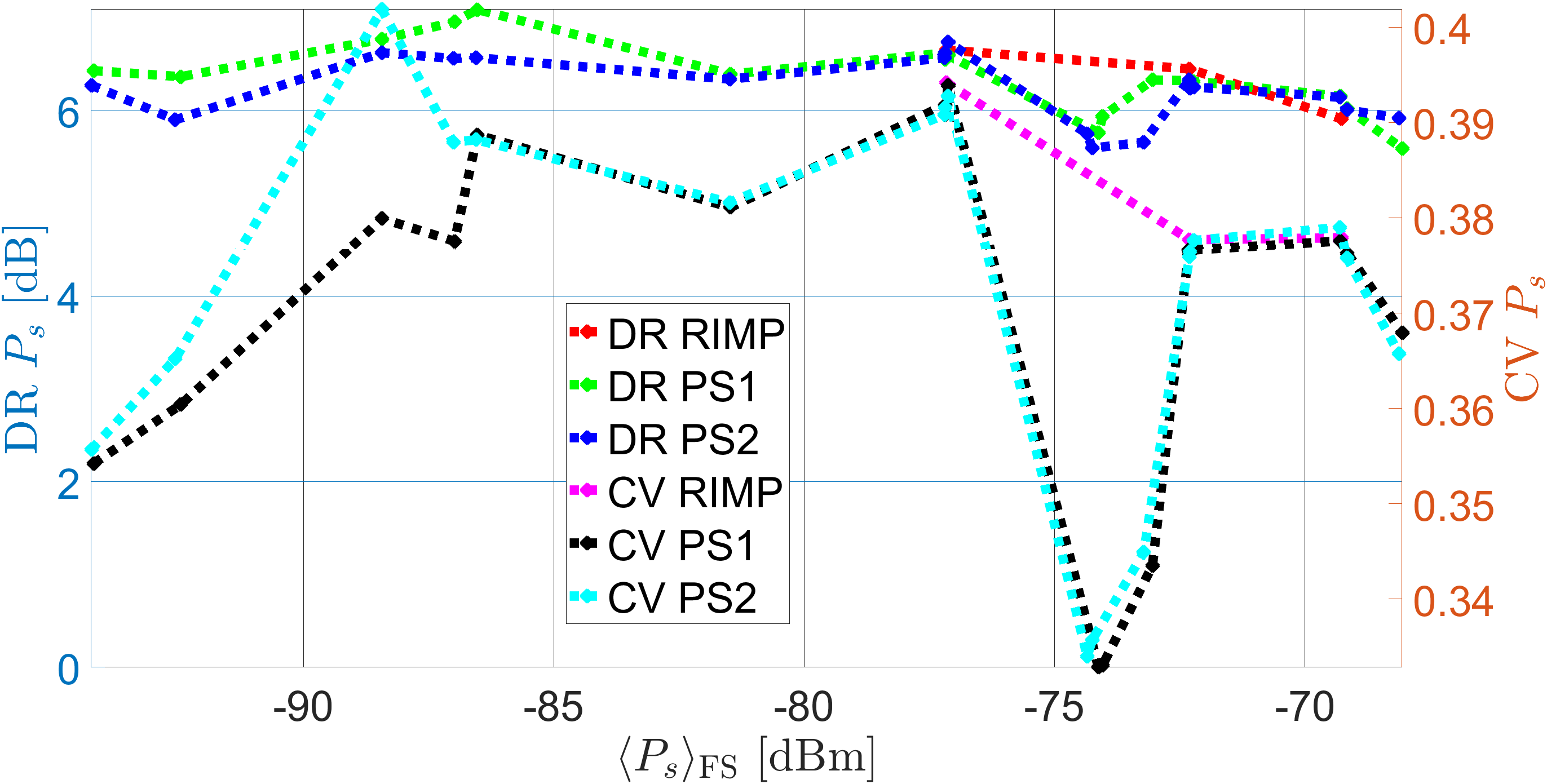}}
\hfil
\subfloat[$P_\mathrm{d}$][\label{F5d}]{\includegraphics[width=1\columnwidth]{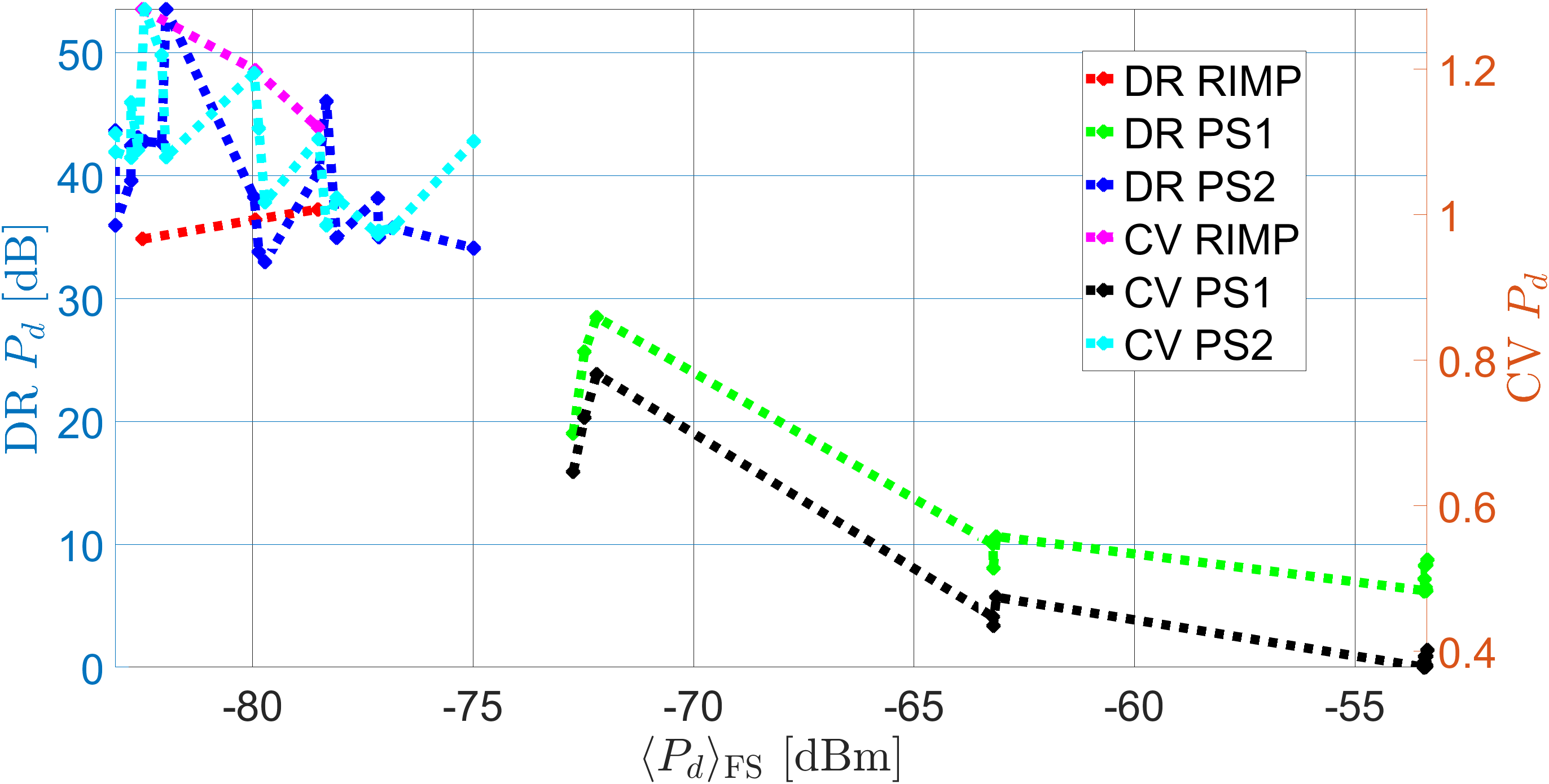}}
\caption{DR and CV of $\Omega$, $K$, $P_\mathrm{s}$ and $P_\mathrm{d}$. Split into cases in which only \ac{RIMP} is excited (red and magenta), cases in which the \ac{CATR} is excited, along or not with \ac{RIMP}, co-polarized with the \ac{RHA} or ``PS1" (green and black), and cases in which the \ac{CATR} is excited, along or not with \ac{RIMP}, cross-polarized with the \ac{RHA} or ``PS2" (blue and cyan). The purpose of these plots is to help identify trends in terms of DR and CV of $\Omega$, $K$, $P_\mathrm{s}$ and $P_\mathrm{d}$ depending on the frequency-averaged values of $\Omega$, $K$, $P_\mathrm{s}$ and $P_\mathrm{d}$, respectively. The split between ``RIMP", ``PS1" and ``PS2" cases is made because they showed different behaviors among them while being similar for the cases they comprise, i.e., ``PS1" cases behave in a relatively consistent manner, which is different to how ``RIMP" and ``PS2" behave.}
\label{F5}
\end{figure*}

\subsection{Received Power}
\label{section: Omega}
For example, for case $1$, where an average $K$-factor of $-9.28$~dB is achieved, the losses from the tip of the 1:4 splitter to the \ac{RHA} are around $56$~dB. Again, this might be suitable for some applications. While for others, such for example, trying to test a device in high \ac{SNR} scenarios, it is not. This and other aspects should be investigated in the future, looking towards the practical challenges of \ac{OTA} testing.
The received power or $\Omega$ [dBm] statistics are presented in columns $5-7$ in Table~\ref{T1}. It can be observed that the frequency-averaged received power $\langle \Omega\rangle_\mathrm{FS}$ is higher for cases in which the \ac{CATR} is excited in co-polarization with the \ac{RHA} ("PS1", or cases $4-6$ and $10-24$). In fact, in the ``PS1" cases without attenuators in the \ac{CATR} branch, the $\langle \Omega\rangle_\mathrm{FS}$ remains almost unchanged. This is because the direct-\ac{LOS} component is dominating, which can be observed in the high values of $\langle K\rangle_\mathrm{FS}$ (column $2$), which results in a relatively strong $\langle P_\mathrm{d} \rangle_\mathrm{FS}$ (column $11$) and relatively weak $\langle P_\mathrm{s} \rangle_\mathrm{FS}$ (column $8$). Therefore, adding a \ac{RIMP} component by exciting the \ac{RIMP} port or changing the absorber configuration does not have a noticeable effect on $\langle \Omega\rangle_\mathrm{FS}$. 
The system losses of the \ac{CATR} system when no polarization discrimination is applied, i.e. when there is co-polarization ("PS1"), are much lower than those of the \ac{RC} excitation ("R"), which was expected, since usually a \ac{LOS} scenario is going to be less lossy than a \ac{NLOS} one. 
On the other hand, ``PS2" cases have significantly lower $\langle \Omega\rangle_\mathrm{FS}$ than their matching ``PS1" cases and, in addition, the ``C" ``PS2" cases ($7-9$) have a lower $\langle \Omega\rangle_\mathrm{FS}$ than the ``R" cases ($1-3$) while having the latter a lower $\langle K\rangle_\mathrm{FS}$ (column $2$). 
As for the addition of absorbers, ``PS1" cases without attenuation are almost not affected, while ``PS2" cases are affected more (larger $\langle \Omega\rangle_\mathrm{FS}$ decrease) in the transition from ``NoAs" to ``BAs" than in the transition from ``BAs" to ``AAs". As expected, the back absorber attenuates most of the \ac{CATR} power. Meanwhile, ``R" cases suffer a larger $\langle \Omega\rangle_\mathrm{FS}$ decrease when going from ``BAs" to ``AAs" than when going from ``NoAs" to ``BAs". This is because the losses for the \ac{RIMP} component are impacted by chamber loading, which is increased considerably more in the case of going from ``BAs" to ``AAs" ($1$ absorber to $6$ absorbers) than when going from ``NoAs" to ``BAs" ($0$ absorbers to $1$ absorber).

Now moving to Fig.~\ref{F5a}, it can be observed that below the $\langle \Omega\rangle_\mathrm{FS}$ of case $37$ of $-76$~dB, there is a high increase in the $\Omega$ \ac{CV}, which reaches $1$ for the lowest $\langle \Omega\rangle_\mathrm{FS}$ cases. Above that, there is a mild trend to a decreased $\Omega$ \ac{CV} with higher $\langle \Omega\rangle_\mathrm{FS}$, until reaching a floor at around $0.4$. The behavior of $\Omega$ \ac{CV} resembles an exponential decay with increasing $\langle \Omega\rangle_\mathrm{FS}$ values. It is worth noting that the behavior of $\Omega$ \ac{CV} ``PS1" is different to that of ``PS2" and ``R", needing a higher $\langle \Omega\rangle_\mathrm{FS}$ to achieve a $\Omega$ \ac{CV} as low as the one of ``PS2" and ``R". The same trends apply to $\Omega$ \ac{DR}, being worth noting that there are some high $\langle \Omega\rangle_\mathrm{FS}$ cases of ``PS1" that exhibit different $\Omega$ \ac{DR} for approximately the same $\langle \Omega\rangle_\mathrm{FS}$.

\subsection{K-factor}

The frequency-averaged $K-$factors or $\langle K\rangle_\mathrm{FS}$ results are shown in columns $2-4$ from Table~\ref{T1}. As seen, $\langle K\rangle_\mathrm{FS}$ covers a wide range from $-9.2$ to $40.8$~dB. The two extreme values correspond to cases $1$ (\ac{RC} excitation only and no absorbers) and case $6$ (\ac{CATR} excitation only in co-polarization with the \ac{RHA} and with all absorbers), respectively. One relevant aspect that can be observed is that the cases of either \ac{RC} excitation only and those in which the \ac{CATR} is excited (only or along with \ac{RIMP}) in the cross-polarization of the \ac{RHA} ("PS2"), generally yield low $\langle K\rangle_\mathrm{FS}$. Conversely, the cases in which the \ac{CATR} is excited (only or along with \ac{RIMP}) in the co-polarization of the \ac{RHA} ("PS1"), yield higher $\langle K\rangle_\mathrm{FS}$. The intervals of $\langle K\rangle_\mathrm{FS}$ produced by the \ac{RC} excitation only and \ac{CATR} excitation only in the co-polarization of the \ac{RHA} ("PS1"), for each of the absorber configurations, are $[-9.2, 20.7]$~dB for the case with no absorbers ("NoAs"), $[-7.6, 35]$~dB for the cases with the back absorber (``BAs"), and $[-5.5, 40.8]$~dB for the cases with all absorbers (``AAs"). These intervals contain all the $\langle K\rangle_\mathrm{FS}$ of the rest of the cases. Moreover, the interval of $\langle K\rangle_\mathrm{FS}$ of the ``BAs" cases is fully overlapped with the intervals of $\langle K\rangle_\mathrm{FS}$ of the ``NoAs" and ``AAs" cases.

In Fig.~\ref{F5b}, it can be observed that the frequency variations heavily depend on the excitation of the system. For \ac{RC} excitation only and all cases in which the \ac{CATR} is excited in cross-polarization with the \ac{RHA} ("PS2"), both the \ac{CV} and the \ac{DR} are far superior to those of the cases where the \ac{CATR} is excited in co-polarization with the \ac{RHA} ("PS1"). It is relevant to note that, for similar $\langle K\rangle_\mathrm{FS}$ values, the use of the co-polarized \ac{CATR} signal provides a much more frequency-stable $K-$factor than using the \ac{CATR} in cross-polarization. As for the \ac{RC} excitation only, it performs, at similar $\langle K\rangle_\mathrm{FS}$ values, quite similarly to the \ac{CATR} in cross-polarization regarding frequency stability. Finally, there is a trend in the case of the co-polarized \ac{CATR} of decreasing frequency variations with increasing $\langle K\rangle_\mathrm{FS}$, although it is not completely monotonic. On the one hand, this behavior is expected because the estimator has a narrower \ac{CI} for higher $\langle K\rangle_\mathrm{FS}$. On the other hand, this behavior might also be explained because of the direct coupling between the \ac{CATR} signal and the \ac{RHA}, which dominates when $\langle K\rangle_\mathrm{FS}$ is high, is more stable with the frequency than the stirred and other unstirred paths which are not the \ac{LOS} itself.

\begin{figure}[!t]
\centering
\includegraphics[width=1\columnwidth]{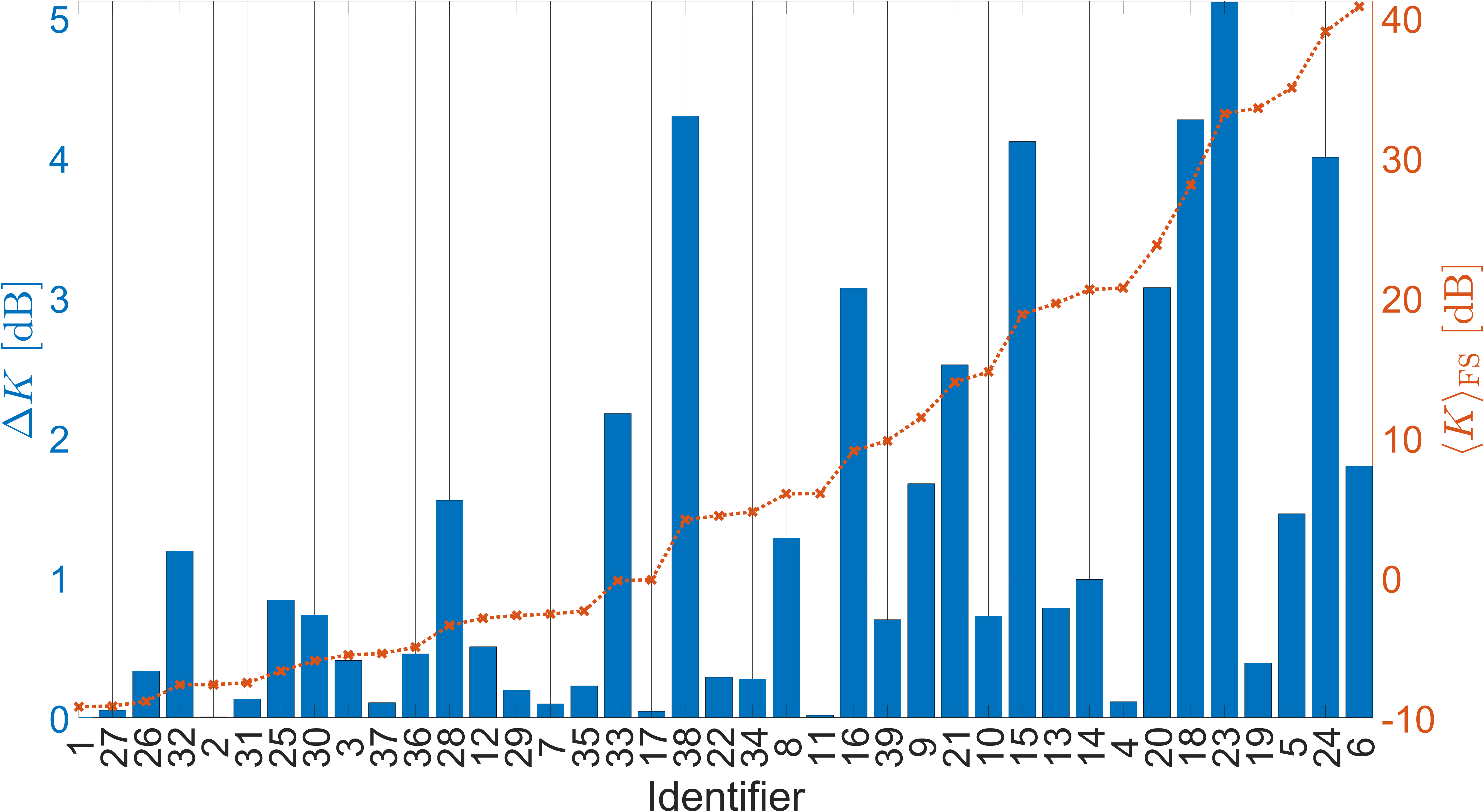}
\caption{Sorted $K-$factors and increments between them.}
\label{F6}
\end{figure}

Fig.~\ref{F6} shows all the cases sorted in increasing order of the frequency-averaged $K-$factor. The $39$ measured cases offer a granularity of the $K-$factor control of $1.3$~dB on average from $-9.2$~dB to $40.8$~dB. However, there are some large increments of up to $5.1$~dB going from case $18$ to case $23$, as shown in Fig.~\ref{F6}. All other transitions are rather smooth. 

\subsection{Stirred Power}

The stirred power or $P_\mathrm{s}$ [dBm] statistics are presented in columns $8-10$ from Table~\ref{T1}. It can be observed that the frequency-averaged stirred power $\langle P_\mathrm{s} \rangle_\mathrm{FS}$ is higher for the cases where no absorbers are used, decreasing as more absorbers are added. It must be noted that all ``PS1" and ``PS2" cases, including the ones with \ac{CATR} excitation only, behave in a very similar manner in terms of $\langle P_\mathrm{s} \rangle_\mathrm{FS}$, $P_\mathrm{s}$ \ac{CV} and $P_\mathrm{s}$ \ac{DR} (see Fig.~\ref{F5c}). This is relevant, since it indicates that the stirred components produced by the \ac{CATR} mostly impinge in the \ac{RHA} with polarization balance or, at least, that the stirrers make the polarizations impinge similarly the \ac{RHA} on average when the \ac{RHA} is in that orientation. Furthermore, in the ``RaC" cases without \ac{RC} branch attenuators, $P_\mathrm{s}$ is dominated by the contribution from the \ac{RC} excitation. Moreover, the ``C" cases and the ``RaC" ``x0ATR" cases suffer very large decreases in $\langle P_\mathrm{s} \rangle_\mathrm{FS}$ when the back absorber is placed, being the decrease milder when adding the rest of the absorbers. This, as already shown in Section~\ref{section: Omega}, implies that most of the \ac{CATR} power which is not directly coupled with the \ac{RHA} is captured by the back absorber. For the ``R" cases, it works the other way around due to the larger increase of chamber load when adding the rest of the absorbers compared to adding just the back absorber.
It is also worth noting that, for some of the ``AAs" cases, in particular, $6$, $9$, $24$, and $39$, the $\langle P_\mathrm{s} \rangle_\mathrm{FS}$ is very close to the average $\langle NL \rangle_{\mathrm{FS}}[\mathrm{dBm}]$ of $-96.3$~dBm. This could impact the estimated $K-$factor for those cases, since the actual $\langle P_\mathrm{s} \rangle_\mathrm{FS}$ might be lower than what is shown here, but the noise floor of the \ac{VNA} limits how low it can be. Finally, there are not any identifiable trends in $P_\mathrm{s}$ \ac{CV} and $P_\mathrm{s}$ \ac{DR} for $\langle P_\mathrm{s} \rangle_\mathrm{FS}$, aside from having a relatively low range of \ac{CV} and \ac{DR} values (see the scales of \ac{CV} and \ac{DR} for the rest of plots from Fig.~\ref{F5}), i.e., all considered cases perform in a more similar way than in the case of the rest of analyzed parameters.

\subsection{Unstirred Power}

The unstirred power or $P_\mathrm{d}$ [dBm] statistics are presented in columns $11-13$ in Table~\ref{T1}. The frequency-averaged stirred power $\langle P_\mathrm{d} \rangle_\mathrm{FS}$ is higher for the ``PS1" cases, where the $P_\mathrm{d}$ contribution is dominated by the direct coupling between the \ac{RHA} and the \ac{CATR}, since no relevant changes in $\langle P_\mathrm{d} \rangle_\mathrm{FS}$ are observed when changing the absorber configurations, which has an impact on the unstirred components that are not the direct coupling. This is observed even for the ``20ATC" cases. However, in these some differences start to appear when changing absorber configurations. For ``PS2" cases, either for ``RaC" ``x0ATR" and ``C" cases, it is observed that the reduction in $\langle P_\mathrm{d} \rangle_\mathrm{FS}$ when going from ``NoAs" to ``BAs" configurations is much larger (around $5$ dB) than when going from ``BAs" to ``AAs". This means that, in these ``NoAs" cases, the contribution to $P_\mathrm{d}$ through the \ac{CATR} excitation in cross-pol to the \ac{RHA} is dominated by the unstirred multipath components, which are strongly attenuated when the back absorber is placed.

From Fig.~\ref{F5d}, it can be highlighted that, unlike for the rest of the analyzed parameters, there is a gap between the achieved $\langle P_\mathrm{d} \rangle_\mathrm{FS}$ values of ``PS1" cases, and those from ``PS2" and ``R" cases. This is due to the strong coupling between the \ac{CATR} and the \ac{RHA} for ``PS1" cases, which, even with 20 dB attenuators provide larger $\langle P_\mathrm{d} \rangle_\mathrm{FS}$ values than any of the ``PS2" and ``R" cases. As for $P_\mathrm{d}$ \ac{CV} and $P_\mathrm{d}$ \ac{DR}, there is some tendency to decrease for both of them with increasing values of $\langle P_\mathrm{d} \rangle_\mathrm{FS}$.

\subsection{Goodness-of-fit test}
\label{section: GoFres}

\begin{figure}[!t]
\centering
\includegraphics[width=1\columnwidth]{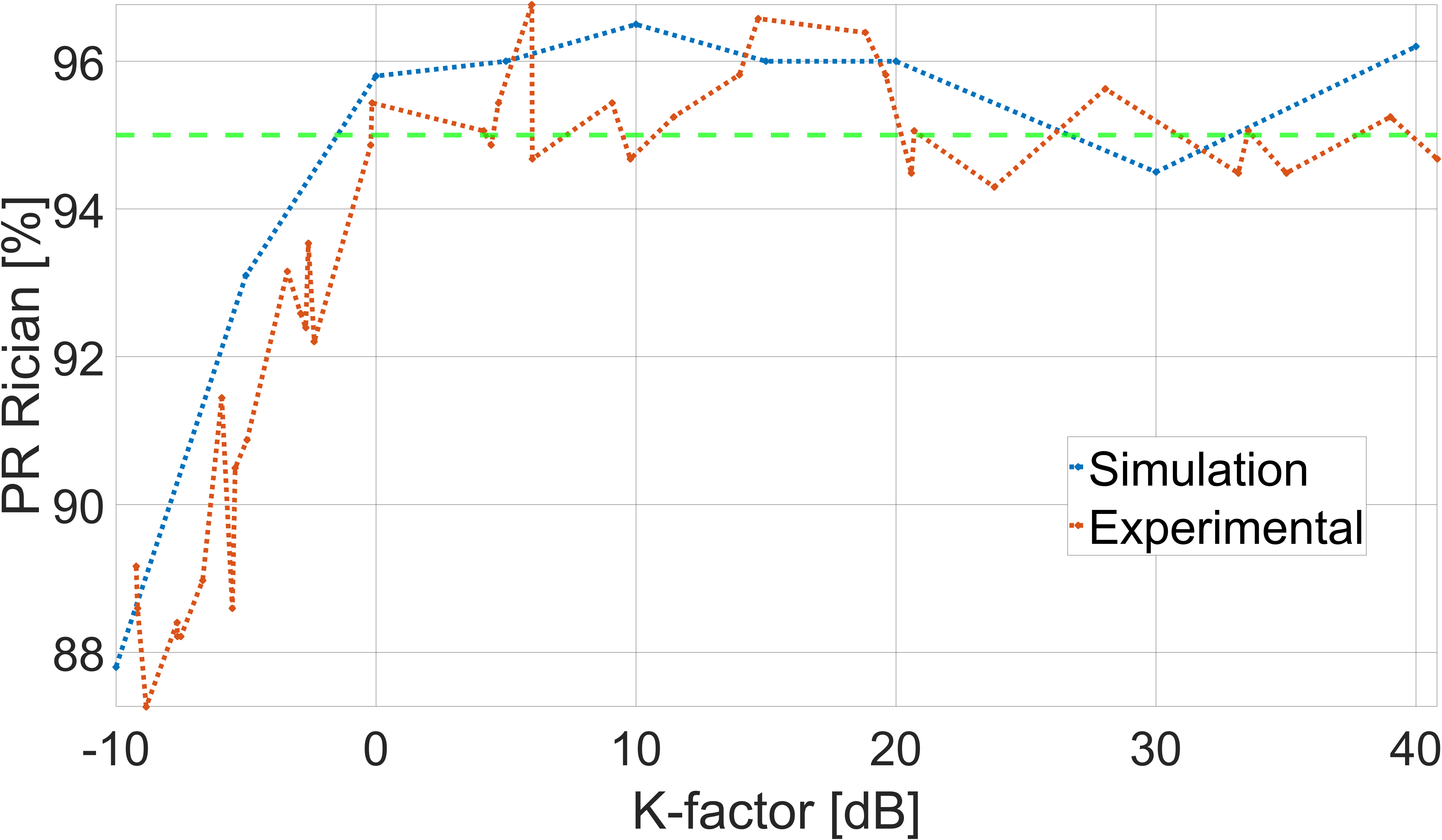}
\caption{Bootstrap \ac{AD} \ac{GoF} test for Rician distribution versus $K-$factor from measurements and simulations. The green line is set at the expected $95\%$ PR. Note that the $K-$factor in the x-axis corresponds to $\langle K\rangle_\mathrm{FS}$ for the experimental results.}
\label{F7}
\end{figure}

The \ac{GoF} tests at each \ac{FS} of every of the $39$ cases are applied to the $600$ samples, which, as we already established in Section~\ref{section: indep}, are independent for all cases.

In Table~\ref{T1}, the \ac{PR} of the bootstrap-based \ac{AD} \ac{GoF} tests with $\alpha$ set to $0.05$ is displayed in columns $14$ and $15$. This \ac{PR} is defined as the number of \ac{FS} where the \ac{GoF} test could not reject the null hypothesis $H_0$ divided by the total number of \ac{FS}. $H_0$ is the data that comes from a Rician distribution in one case; in the other, $H_0$ is the data that comes from a Rayleigh distribution.

Since $\alpha$ is set to $0.05$, the expected \ac{PR} if the data comes from a distribution of the family of $H_0$, i.e., if $H_0$ is true, would be $95\%$. Therefore, the expected \ac{PR} for the \ac{GoF} test with $H_0$ being that the data comes from a Rician distribution, is $95\%$, since we expect to get a signal that is Rician distributed in all cases. However, the results are not completely aligned with the expectations. In some cases, we get close to the expected $95\%$ \ac{PR}, but in other cases, we get lower ones, being the minimum $87.3\%$. In particular, it can be observed that there appears to be some correlation between lower $\langle K\rangle_\mathrm{FS}$ and lower \acp{PR} for Rician distribution as the $H_0$. This can be observed in Fig.~\ref{F7}, where simulations have also been included to confirm if the observed behavior of the experimental data is expected or not. The simulations consisted of generating $1000$ random samples with a size of $600$ samples from a Rician distribution with a fixed $K-$factor. Then, the same bootstrap-based \ac{AD} \ac{GoF} test for $H_0$ being that the distribution of the data is Rician was applied to each of the $1000$ sets of $600$ samples, obtaining a $1000$ rejections or not of $H_0$. From that, the \ac{PR} was computed as the percentage of not rejection of $H_0$. This was repeated for as many fixed $K-$factors as points are in the simulations curve ($9$). Although the underlying reasons for the observed behavior of the \ac{PR} of the \ac{GoF} test w.r.t. the $K-$factor should be further studied and tackled, which we consider falling beyond the scope of this work, the fact that we observe very similar behavior for the experimental data leads us to the conclusion that the experimental data for each of the considered cases follows a Rician distribution. Thus, the consideration of $K-$factor, $\Omega$, $P_{\mathrm{s}}$, and $P_{\mathrm{d}}$ as fundamental parameters to describe our experimental data is appropriate.

On the other hand, we observe an inverse relation between $K-$factor and \ac{PR} for the \ac{AD} \ac{GoF} test with $H_0$ being that the data comes from a Rayleigh distribution. This is something to be expected, i.e., a Rayleigh distribution is a Rician distribution with a $K-$factor equal to $0$ or $-\infty$~dB, so the lower the $K-$factor, the closer the \ac{PR} should be to $95\%$. These results show that the considered cases are not achieving a pure Rayleigh distributed signal, although some of them, especially case $1$, come close to it. In particular, cases $1$, $2$, $26$, $27$, $31$ and $32$ have a \ac{PR} for Rayleigh very close to or over $90\%$, with a maximum $\langle K\rangle_\mathrm{FS}$ of $-7.6$~dB. Therefore, although accepting some error, we could approximate cases with $\langle K\rangle_\mathrm{FS}$ lower than $-7.6$~dB by a Rayleigh model. On the other hand, the increased rejection (lower \ac{PR}) of $H_0$ of Rayleigh distribution when $\langle K\rangle_\mathrm{FS}$ becomes larger, i.e., the distribution becomes less similar to a Rayleigh, indicates that the applied \ac{GoF} test is working properly.

On another note, the fact that the lowest achieved $\langle K\rangle_\mathrm{FS}$ is $-9.2$~dB (case $1$) is because turntable stirring is not being used. As shown in \cite{StirUnstir}, this method effectively reduces $\langle K\rangle_\mathrm{FS}$. A measurement in the same setup as case $1$ but with turntable stirring was performed, obtaining a $\langle K\rangle_\mathrm{FS}$ of $-17.6$~dB, thus obtaining a channel much closer to \ac{RIMP}. Traditional \ac{RC} measurements such as \ac{TRP} or antenna efficiency are usually performed with turntable stirring in Bluetest's systems.

\subsection{Frequency dependence of the estimated distribution parameters}
\label{section: freqresp}

\begin{table}
\centering
\caption{Fitting parameters and $R^2$ of $K$, $\Omega$, $P_\mathrm{s}$, and $P_\mathrm{d}$.}
\label{T2}
\includegraphics[width=1\columnwidth]{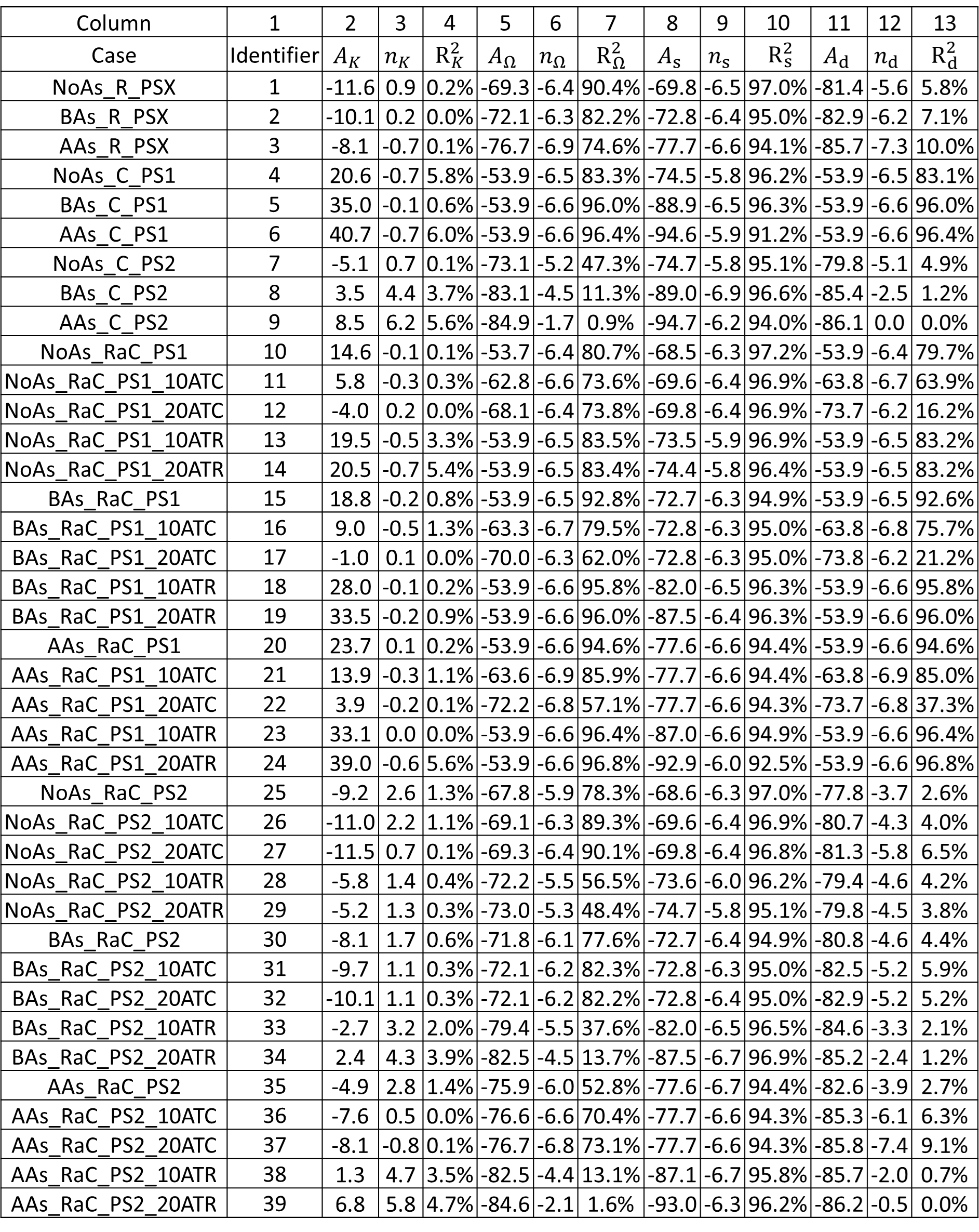}
\end{table}

\begin{figure*}
\centering
\subfloat[$K$][\label{F8a}]{\includegraphics[width=0.66\columnwidth]{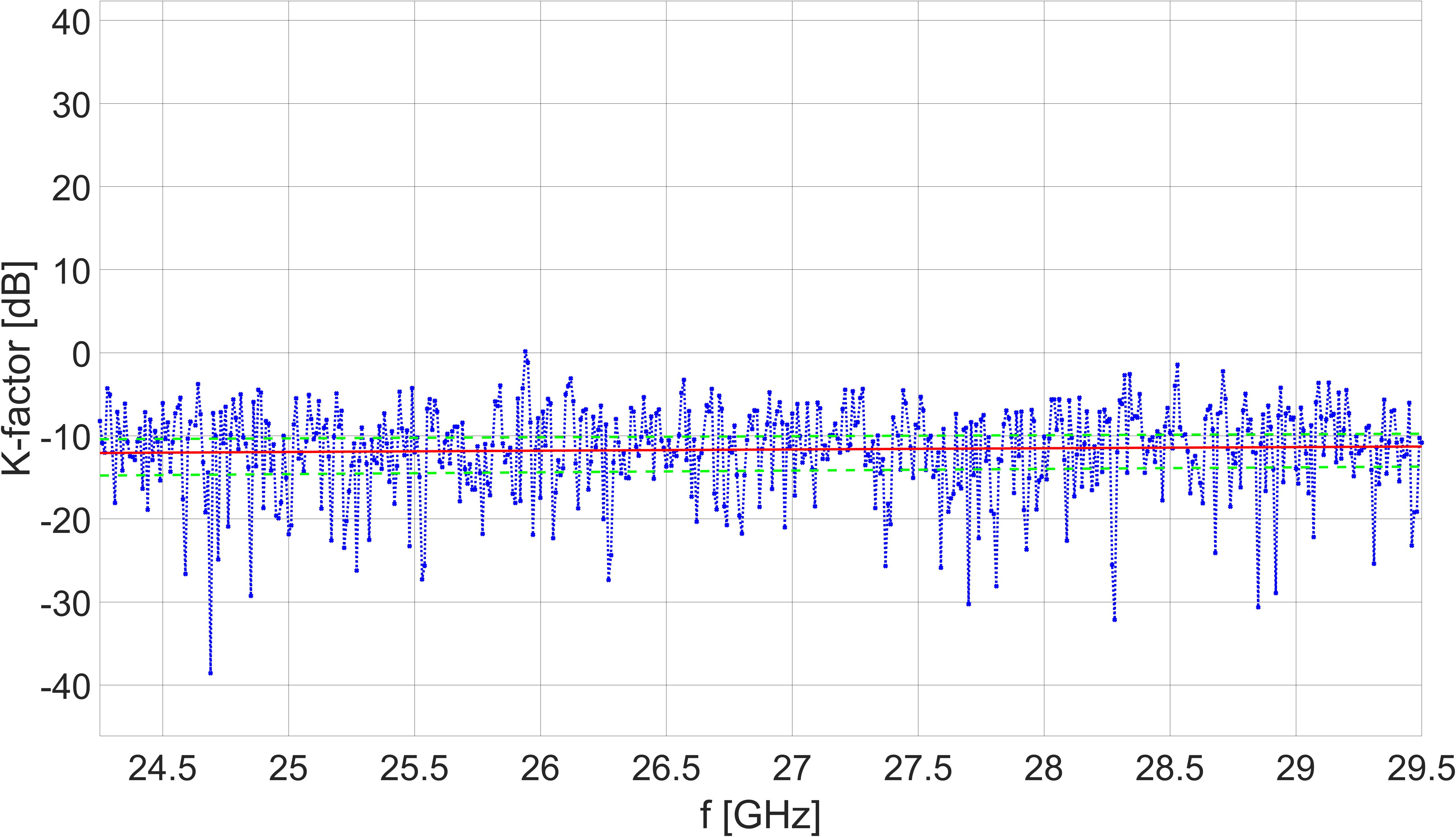}}
\hfil
\subfloat[$K$][\label{F8b}]{\includegraphics[width=0.66\columnwidth]{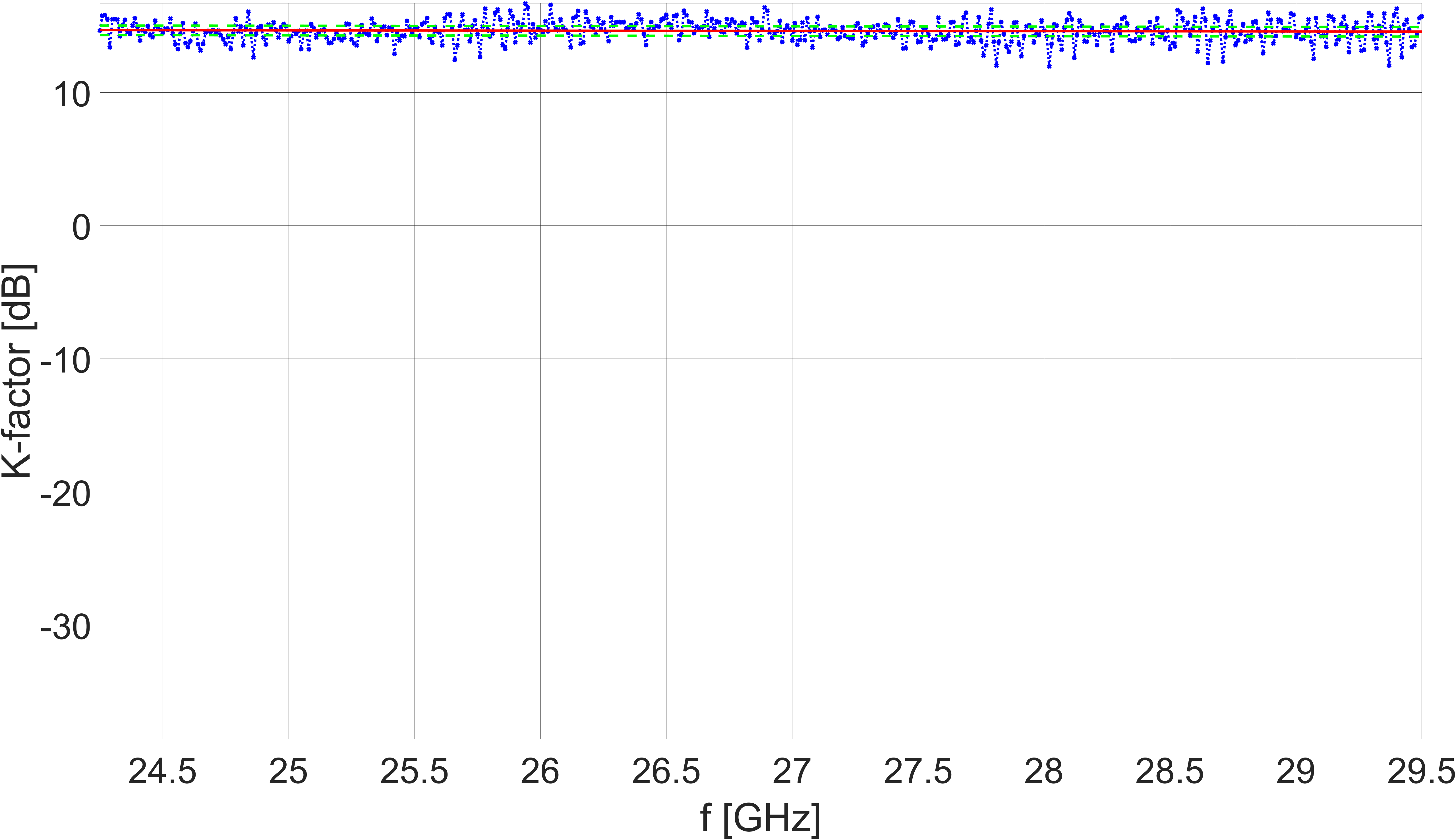}}
\hfil
\subfloat[$K$][\label{F8c}]{\includegraphics[width=0.66\columnwidth]{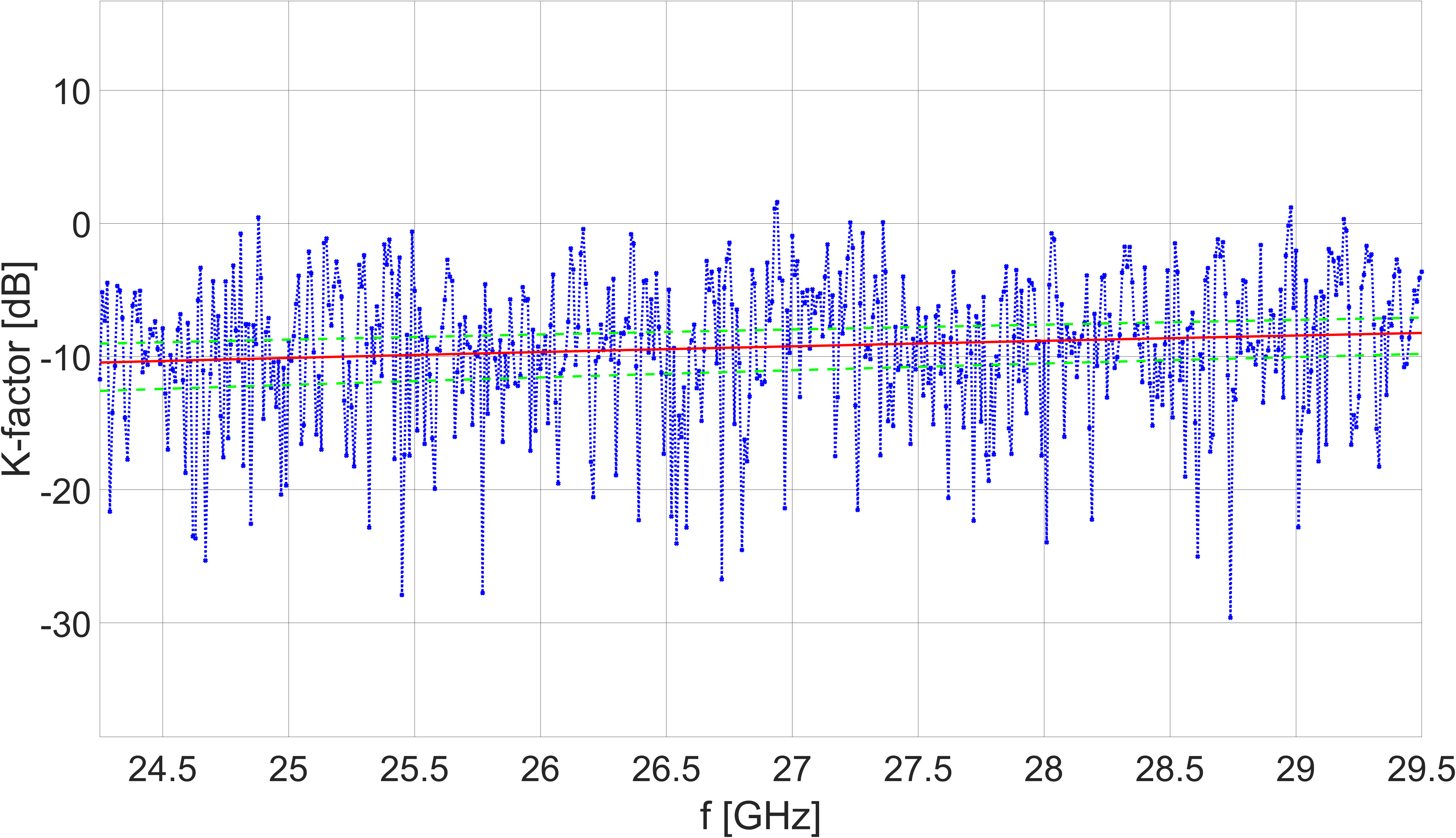}}
\hfil
\subfloat[$\Omega$][\label{F8d}]{\includegraphics[width=0.66\columnwidth]{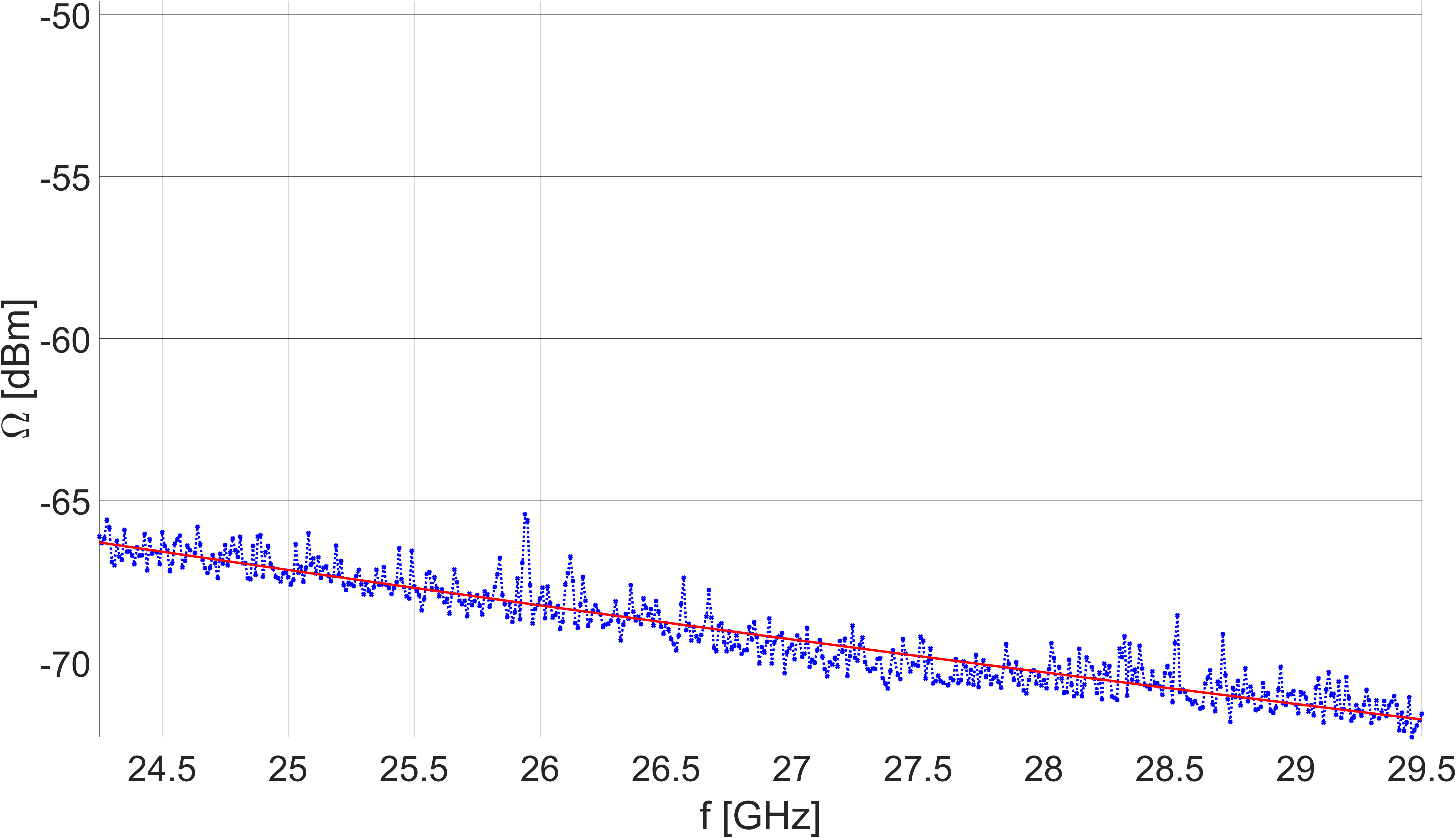}}
\hfil
\subfloat[$\Omega$][\label{F8e}]{\includegraphics[width=0.66\columnwidth]{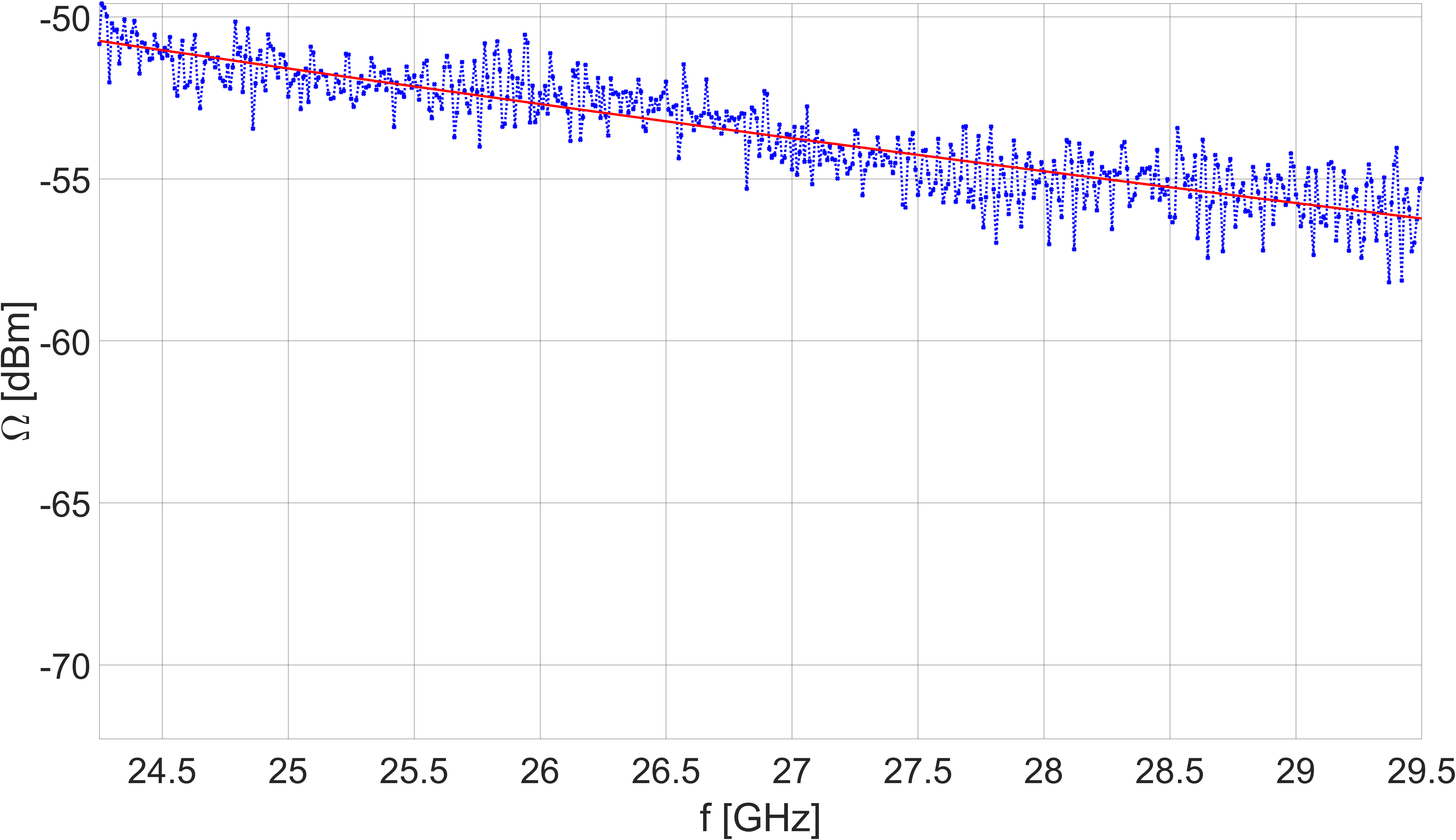}}
\hfil
\subfloat[$\Omega$][\label{F8f}]{\includegraphics[width=0.66\columnwidth]{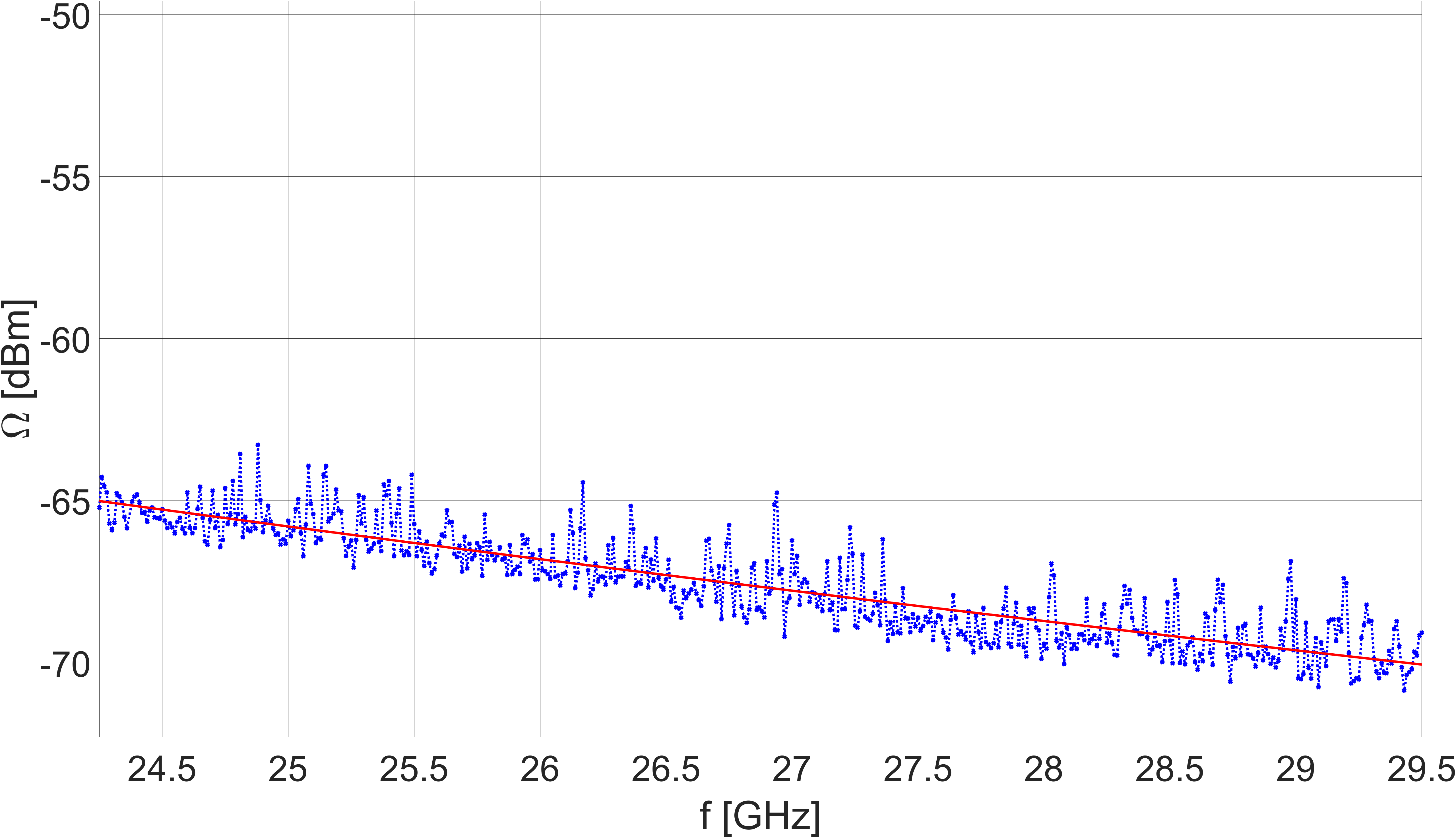}}
\hfil
\subfloat[$P_\mathrm{s}$][\label{F8g}]{\includegraphics[width=0.66\columnwidth]{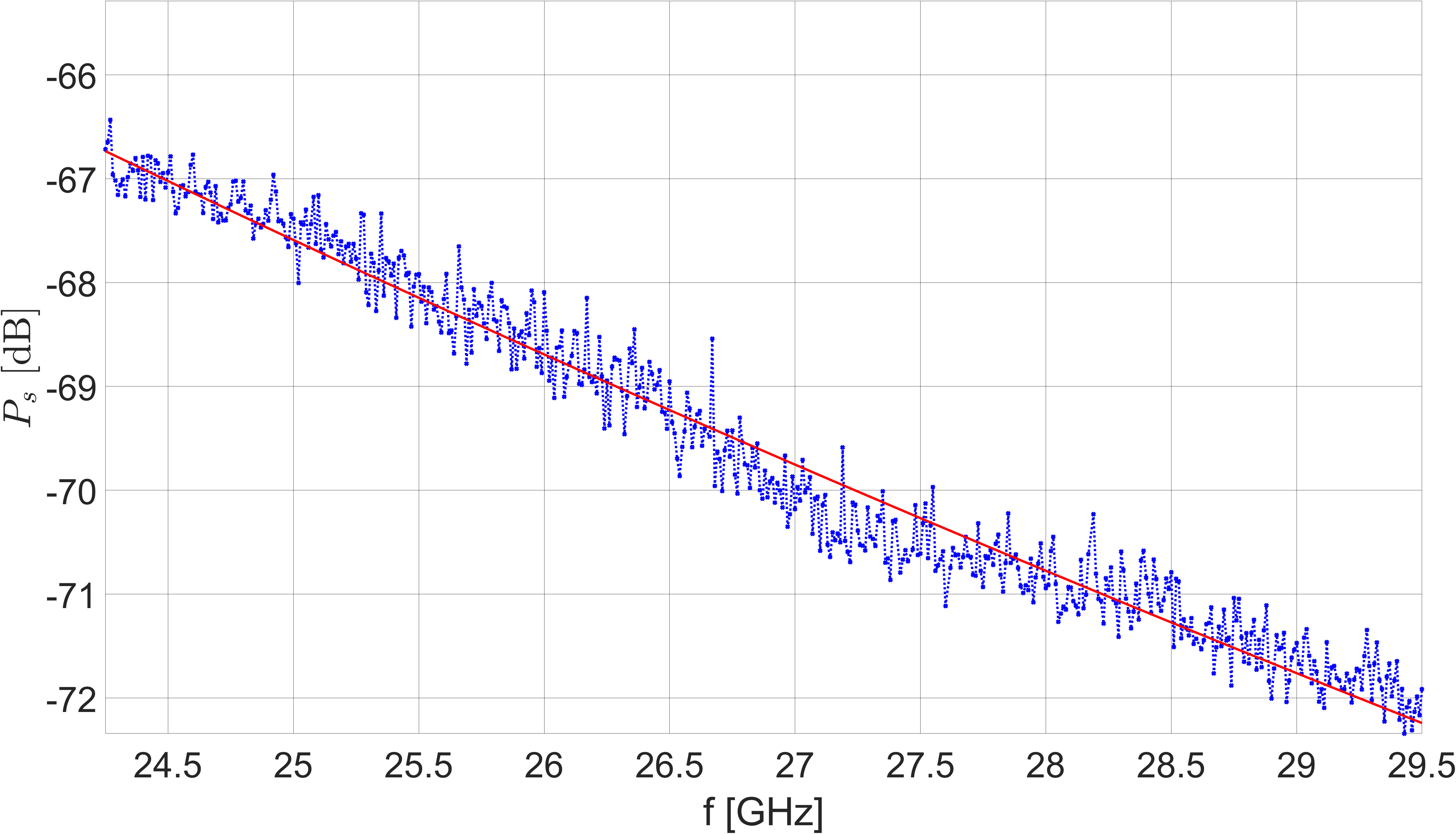}}
\hfil
\subfloat[$P_\mathrm{s}$][\label{F8h}]{\includegraphics[width=0.66\columnwidth]{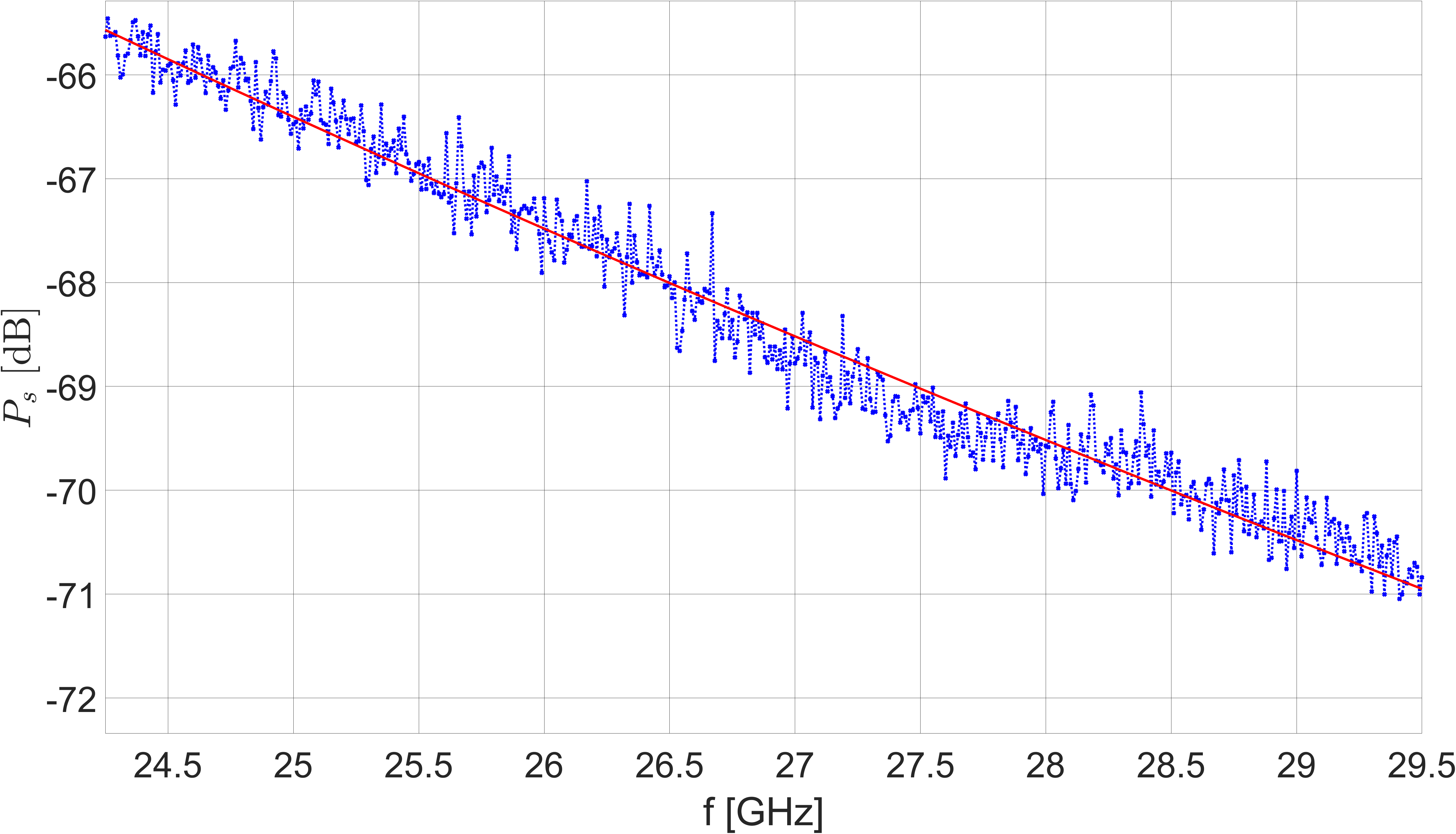}}
\hfil
\subfloat[$P_\mathrm{s}$][\label{F8i}]{\includegraphics[width=0.66\columnwidth]{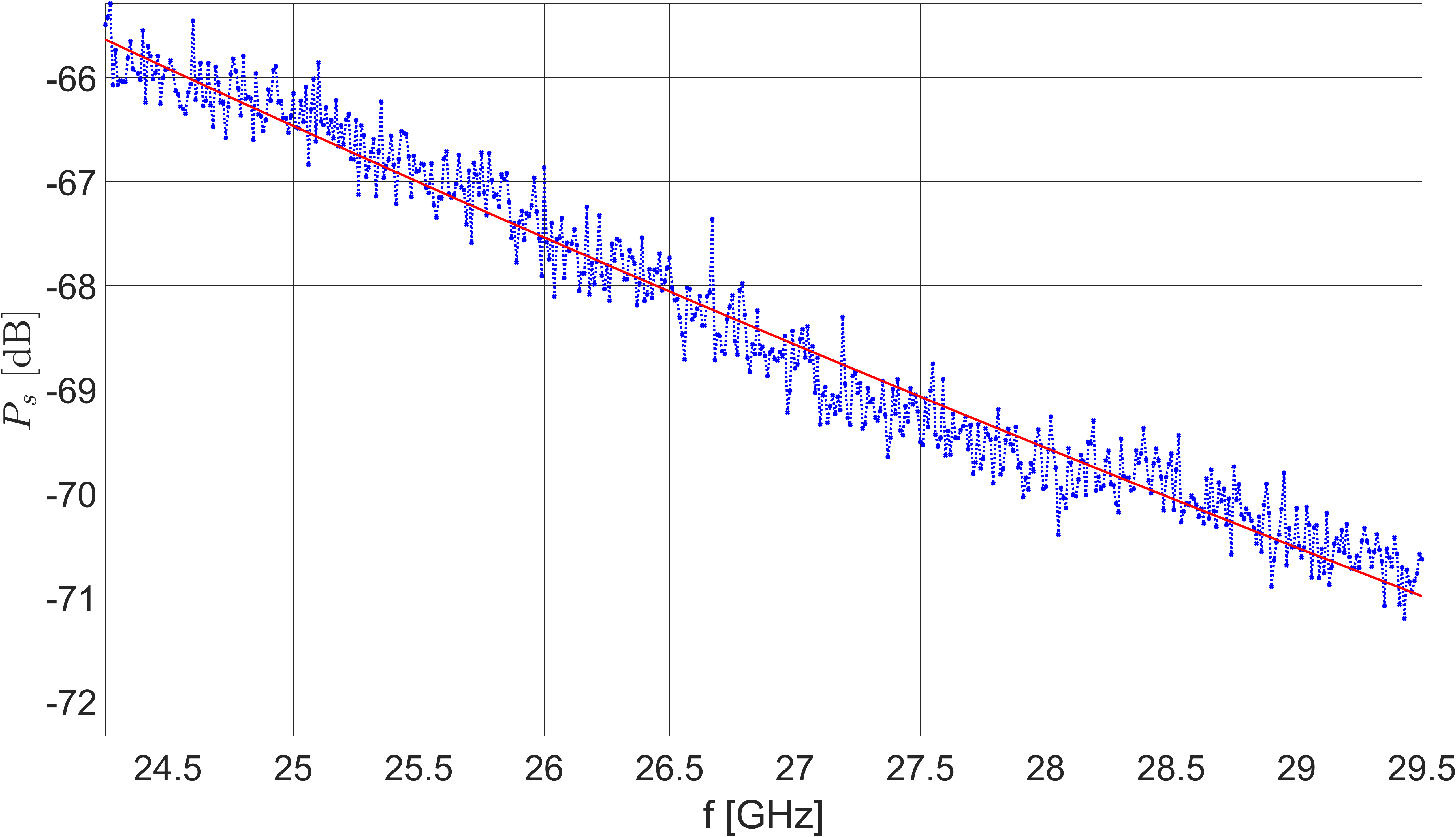}}
\hfil
\subfloat[$P_\mathrm{d}$][\label{F8j}]{\includegraphics[width=0.66\columnwidth]{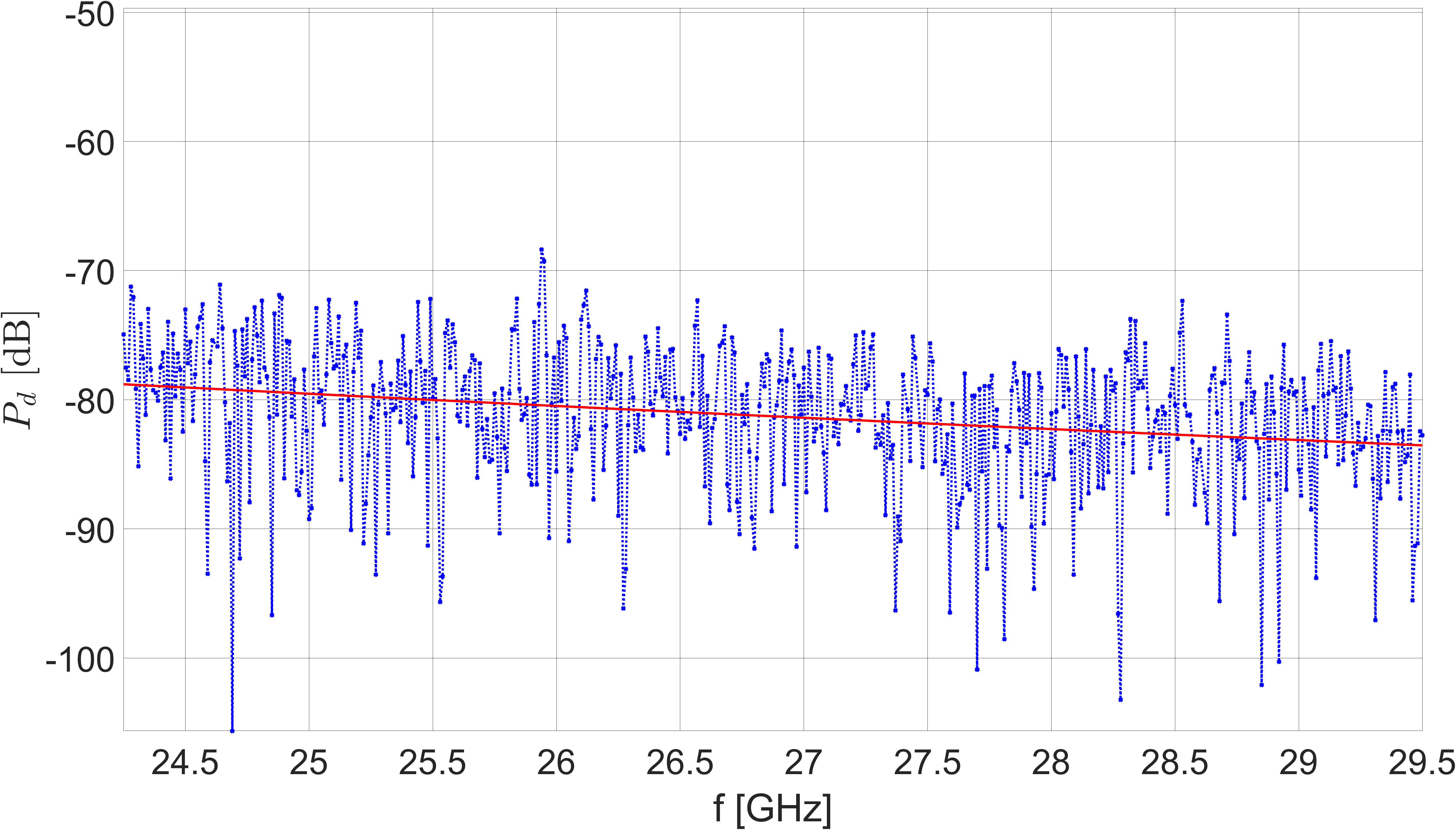}}
\hfil
\subfloat[$P_\mathrm{d}$][\label{F8k}]{\includegraphics[width=0.66\columnwidth]{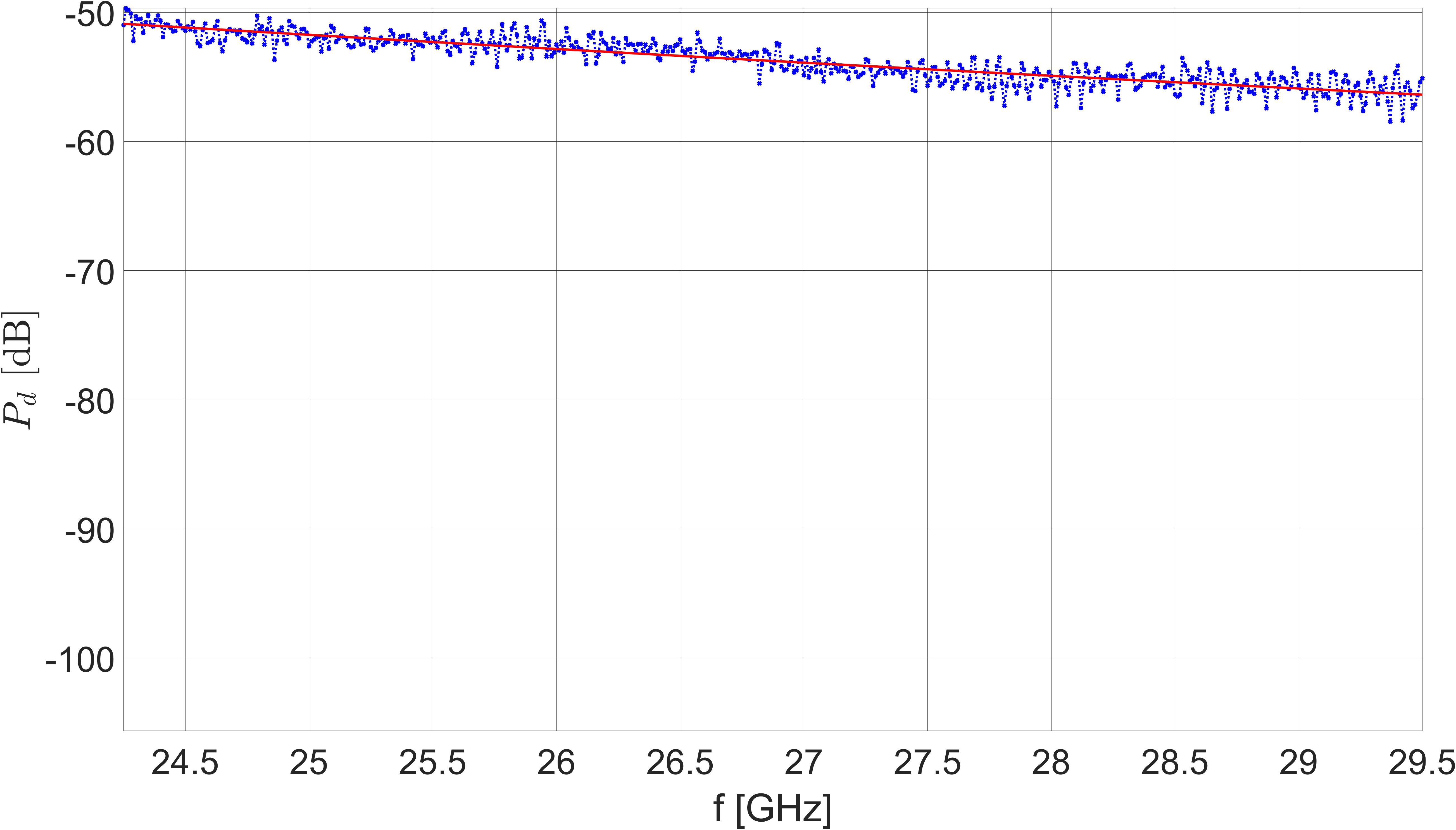}}
\hfil
\subfloat[$P_\mathrm{d}$][\label{F8l}]{\includegraphics[width=0.66\columnwidth]{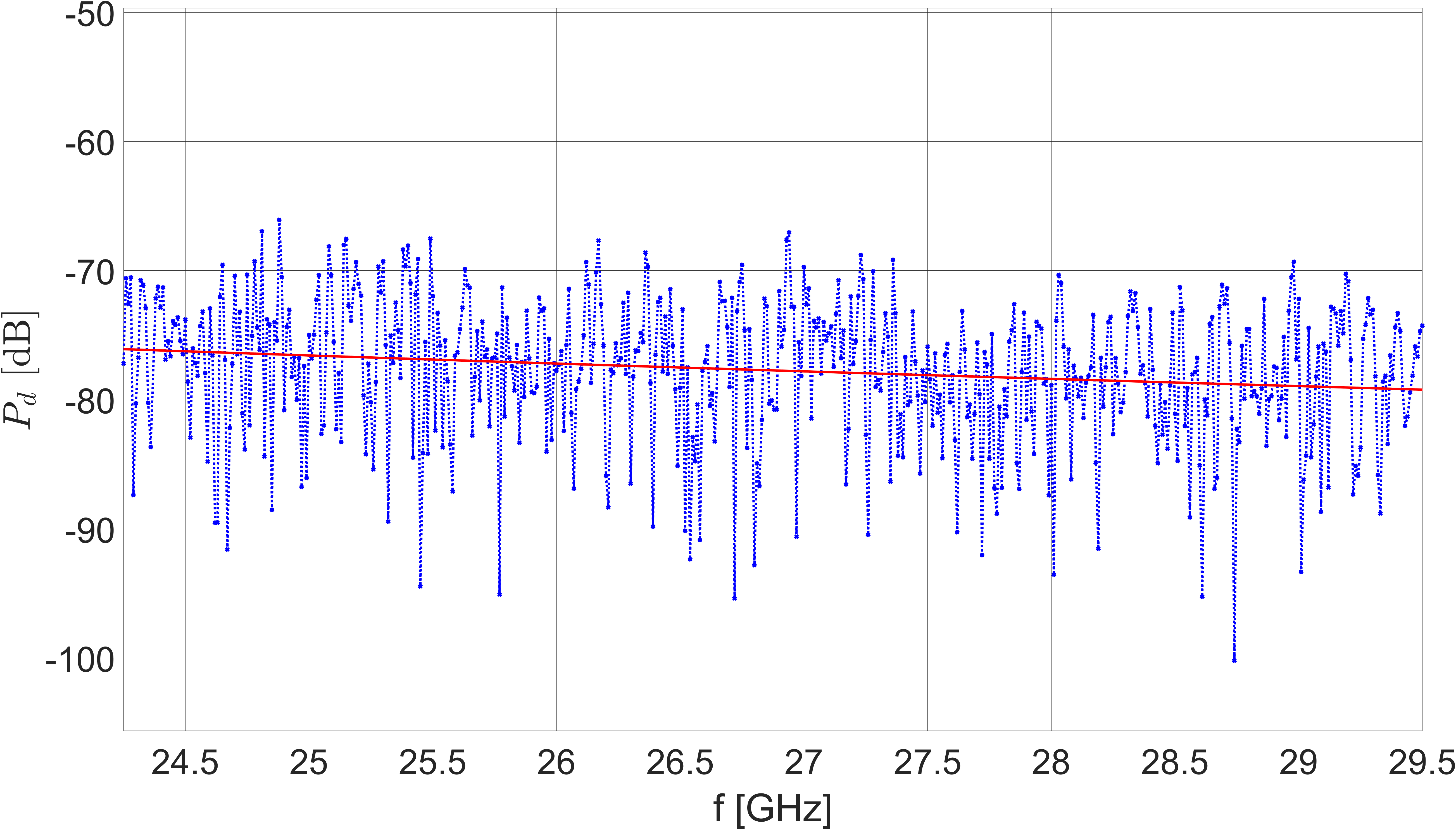}}
\hfil
\caption{Frequency plots of $K$, $\Omega$, $P_\mathrm{s}$ and $P_\mathrm{d}$ for cases, from left to right, $1$, $10$, and $25$. The estimated values or each parameter are represented in blue. The values coming from the fittings are represented in red. $K$ plots also have two green lines that indicate the $95\%$ \ac{CI} of the $K$ estimator, computed from (43) and (44) from \cite{KFEmulRician}.}
\label{F8}
\end{figure*}

In the above sections, we studied the statistics of the generated Rician channels as averaged over frequency or considering their statistics over frequency. In this section, we study their frequency dependence. This is done to understand whether there are relevant trends that we should note for channel emulation purposes. For that, we produce the following fitting curves
\begin{eqnarray}
    K[\mathrm{dB}]&=&A_K+10n_Klog10(f/f_0),\\
    \Omega[\mathrm{dBm}]&=&A_{\Omega}+10n_{\Omega}log10(f/f_0),\\
    P_\mathrm{d}[\mathrm{dBm}]&=&A_\mathrm{d}+10n_\mathrm{d}log10(f/f_0),\\
    P_\mathrm{s}[\mathrm{dBm}]&=&A_\mathrm{s}+10n_\mathrm{s}log10(f/f_0),
\end{eqnarray}
where $f_0$ is set to $27$~GHz. The election of $f_0$ is based on the fact that the considered frequency range of $24.25-29.5$~GHz contains the n257 and n258 \ac{5G} FR2 bands, which are commonly referred to as the $28$~GHz and $26$~GHz bands, respectively, due to their center frequencies. Therefore, we have considered $f_0$ to be in the middle of the bands' center frequencies: $27$~GHz. The parameters for the fitting are $A$ and $n$. $A$ will be the value of the fitted curve at $f_0$, while $n$ will describe the frequency dependence of the fitted data. In particular, $n$ represents the exponent of the frequency for $K$, $\Omega$, $P_\mathrm{s}$, and $P_\mathrm{d}$ expressed in linear units. The fitting results are shown in Table~\ref{T2}, including the $R^2$ value. We will focus on $n$ rather than $A$, since we are interested in analyzing the frequency dependence of $K$, $\Omega$, $P_\mathrm{s}$, and $P_\mathrm{d}$. In addition, the frequency plots of $K$, $\Omega$, $P_\mathrm{s}$, and $P_\mathrm{d}$ and the corresponding fitting curves for cases $1$, $10$, and $25$ can be found in Fig.~\ref{F8}. The selection of cases was made to illustrate three different relevant configurations under the same absorber configuration, having \ac{RC} excitation only in case $1$, \ac{RC} and \ac{CATR} in co-polarization with the \ac{RHA} in case $10$, and \ac{RC} and \ac{CATR} in cross-polarization with the \ac{RHA} in case $25$.

From the presented results, it can be highlighted that both $P_\mathrm{s}$ and $P_\mathrm{d}$ are inversely related to the frequency for all cases (except case $9$, which has no relation between $P_\mathrm{d}$ and frequency). This behavior is also observed in \cite{RC_Rician}. The $n_\mathrm{s}$ values are quite similar for all cases. On the other hand, the $n_\mathrm{d}$ values have a larger variation, being very similar among ``PS1" cases. Since $\Omega$ is the sum of $P_\mathrm{s}$ and $P_\mathrm{d}$, then it is also inversely related with the frequency in all cases. Although this is not plotted in \cite{RC_Rician}, it is also the case there due to $P_\mathrm{s}$ and $P_\mathrm{d}$ being inversely related with frequency, so being $\Omega$ their sum, it is then also inversely proportional to the frequency. The inverse relation of $\Omega$, $P_\mathrm{s}$, and $P_\mathrm{d}$ with the frequency can be observed in Fig.~\ref{F8}~(d-l). In addition, the $n_{\Omega}$ values exhibit the same trend of being stable among ``PS1" cases. Conversely, the $K-$factor does not have a general direct or inverse relation with frequency. It is worth noting that most ``PS1" cases have a negative and small $n_K$, i.e., a slight inverse relation with the frequency (Fig.~\ref{F8b}). In comparison, \cite{RC_Rician} shows a direct dependence on the frequency of $K$-factor when both horn antennas are co-polarized, which would be comparable to the ``PS1" cases. In our work, as stated above, these cases mostly show a slight inverse relation with frequency. This implies that the inverse relation with the frequency of $P_\mathrm{d}$ is larger (more negative) than the one from $P_\mathrm{s}$ in our work, i.e. $n_d<n_s$. In \cite{RC_Rician}, it happens that $n_d>n_s$. While we do not go into the reasons why this is different in both works or, more generally, into the frequency dependence modeling of $K$, $\Omega$, $P_\mathrm{s}$, and $P_\mathrm{d}$, we acknowledge it as a point of future work. On the other hand, most ``PS2" cases have a positive and generally larger $n_K$, i.e., a direct relation with the frequency (Fig.~\ref{F8c}). ``R" cases show a direct relation with frequency for the ``NoAs" (Fig.~\ref{F8a}) and ``BAs" cases, but inverse for the ``AAs" case. In comparison, we have \cite{Kildal_RC_KF_formula}, where a similar \ac{RC} (RTS chamber) is used in its traditional, no Rician emulation, mode and with different loadings. As can be seen, the $K$-factor is inversely related to the frequency in all loading conditions, although this inverse relationship seems to weaken as the frequency increases. In our work, we observe a direct relation with frequency for less loaded setups (``NoAs" and ``BAs") and an inverse relation for the heavily loaded scenario (``AAs"). Although we do not focus on the comparison with \cite{Kildal_RC_KF_formula}, we say that the differences in the behavior of $K$ might be explained by the large difference in considered frequencies, which, for example, change which component dominates the quality factor of the chamber $Q$, as stated in \cite{Hill_eq}.

On the other hand, the $R^2$ value for $P_\mathrm{s}$ is very high for all cases, which implies that the proposed fitting curves explain almost all the variability of $P_\mathrm{s}$ with the frequency. In the case of $P_\mathrm{d}$, the $R^2$ values are high for the ``PS1" cases and low for the ``R" and ``PS2" ones. This is because ``PS1" cases (Fig.~\ref{F8k}) have a $P_\mathrm{d}$ mostly contributed by the direct coupling of the \ac{RHA} and the \ac{CATR}, while in the ``R" and ``PS2" cases, $P_\mathrm{d}$ is mostly contributed by a sum of reflections that do not interact with the stirrers. Such a sum can have large variations due to phase changes of different reflections' path lengths, resulting in constructive or destructive interference. Therefore, it is somewhat expected to observe variations that cannot be approximated by a smooth fitting curve, being this reflected in the much lower $R^2$ values for the fitting of $P_\mathrm{d}$ for ``R" (Fig.~\ref{F8j}) and ``PS2" (Fig.~\ref{F8l}) cases. Since, again, $\Omega$ is the sum of $P_\mathrm{s}$ and $P_\mathrm{d}$, the $R^2$ values of $\Omega$ are a result of a weighted combination of the variability of $P_\mathrm{s}$ and $P_\mathrm{d}$ that can and cannot be explained by the proposed models. Therefore, since all $P_\mathrm{s}$ have high $R^2$ values, the limiting factor is $P_\mathrm{d}$. Hence, the $R^2$ values of $\Omega$ are high when either $P_\mathrm{s}$ is larger than $P_\mathrm{d}$ (low $K$) or when the case is a ``PS1" one, since the $R^2$ of $P_\mathrm{d}$ will be high. Conversely, the $R^2$ values of $\Omega$ are low when $P_\mathrm{s}$ is smaller than $P_\mathrm{d}$ (high $K$) and, at the same time, the case is a ``PS2" one, since the $R^2$ of $P_\mathrm{d}$ will be low. Finally, the $R^2$ values of $K$ are low in all cases, which implies that most variability of $K$ w.r.t. the frequency cannot be explained by the fitted model.

On another note, in Fig.~\ref{F8}~(a-c), the $95\%$ \ac{CI} of the $K-$factor estimator has been plotted for the fitted model. The goal is to determine whether the observed frequency variations in the $K$-factor are a result of the estimator's uncertainty due to the finite sample size (see Fig.~\ref{F4}), or if these variations reflect actual changes in the $K$-factor with frequency that are not accounted for by the fitted models. Since most points ($\gg5\%$) fall outside the \ac{CI}, it can be inferred there are variations of the actual $K-$factor in frequency not captured by the fitted models which are not due to the uncertainty of the $K-$factor estimator. Although not shown, this is also the case for the rest of the cases.

\section{Conclusions}
\label{section: Conclusions}
This work has proposed a practical method to use hybrid \ac{RC} plus \ac{CATR} \ac{OTA} chambers to emulate Rician channels with controllable $K-$factors at \ac{mmWave}. The focus has been on the lower FR2 bands ($24.25-29.5$~GHz). The main goal has been creating a reference Rician channel to characterize beamforming and directional antennas. Therefore, the channel emulation procedure has employed a \ac{RHA} as a reference. We hope this will pave the way to cost-efficient \ac{OTA} testing of directional \ac{mmWave} devices in channels with different $K-$factors that emulate real use cases.

First, the measured data has been analyzed to ensure it is valid for the procedures conducted in the further analysis of the parameters of the Rician distribution, e.g., the $K-$factor, the average received power, as well as the powers of the deterministic and the random signal components. It was established that all measured samples were independent, and the average power was well above the noise floor, i.e., at least 13 dB on average. However, the \ac{SNR} might not be sufficient, or, more generally, the systems' losses might be too high for some of the cases, depending on the testing instruments used, as well as the use case. For example, if one desires to test throughput at high \acp{SNR} or signal levels, only a subset of the configurations may be capable of providing such signal level, as they are right now. This is not unsolvable, e.g. amplifiers, lower losses cables, and/or a 2:1 splitter could be used, but this should be taken care of if actual \ac{OTA} measurements are performed in this system. 

Second, a bootstrap \ac{AD} \ac{GoF} test was used to confirm that the measurement data most likely follows a Rician distribution in all cases while the Rayleigh distribution is a good model for low $K-$factors, i.e., $K\leq-7.6$~dB. 

Third, it has been proven that a wide range of $K-$factors can be generated with the proposed measurement setups. Employing the 39 different considered configurations, the produced frequency averaged $K$-factors or $\langle K\rangle_\mathrm{FS}$ could be varied from $-9.2$~dB to $40.8$~dB, with a maximum increment of $5.1$~dB, while the average increment was $1.3$~dB. 

Fourth, the parameters that characterize the Rician distribution most likely followed by the measurement data, $K$, $\Omega$, $P_\mathrm{s}$, and $P_\mathrm{d}$ were analyzed. From them, several relevant points could be identified: the back absorber is very effective in attenuating the power of the \ac{CATR} signal that is not directly coupled with the \ac{RHA}. It is needed to use the \ac{CATR} signal co-polarized with the \ac{RHA}, which results in high $K-$factors, to have the lowest system losses, achieving also higher frequency stability for the same $\langle K\rangle_\mathrm{FS}$ compared to the cross-polarization cases. $\Omega$, $P_\mathrm{s}$, and $P_\mathrm{d}$ show an inverse relation with the frequency. $K$ does not show a general direct or inverse relation with the frequency. The proposed models to fit to the measurement data can explain most of the frequency variability of $P_\mathrm{s}$ in all cases, of $\Omega$ in most cases, of $P_\mathrm{d}$ in some cases, failing to explain it for $K$. The frequency variability of $K$ is also proven not to be explained just by the uncertainty of the used estimator due to finite sampling.

Future work may include a \ac{MIMO} setup, since the cases of \ac{RC} and \ac{CATR} have been combined through a splitter/combiner. This might include the simultaneous use of the two available \ac{RC} antennas, as well as the two \ac{CATR} polarizations, which can be accessed via two different ports, allowing up to a $4\time4$ \ac{MIMO} channel. On the other hand, one of the system's limitations, especially for the cases where low $K-$factors are generated, is the large loss that the signal suffers. Also, an uncertainty analysis of the measurements and their impact on other aspects of the study, such as the $K-$ factor or the frequency variations of the $K-$factor can be useful. Moreover, frequency modeling of $K$, $\Omega$, $P_\mathrm{s}$, and $P_\mathrm{d}$ is acknowledged as relevant to better understand the system behavior. In addition, the reasons why the \ac{GoF} test is not giving exactly the expected $95\%$ \ac{PR} for Rician distribution should be further investigated. Finally, a relevant limitation of the proposed setup is that it cannot achieve average $K$-factors below $-9.2$~dB, which limits the minimum uncertainty of classic \ac{RC} measurements such as \ac{TRP}. An approach that can extend the low $K$-factors further down is to investigate the possibility of mounting both the \ac{DUT} and the \ac{CATR} on a turntable to achieve additional stirring of the cavity modes. This, in turn, would require an assessment of whether or not $600$ samples would be enough to estimate lower $K$-factors or even if using another estimator might be a better option for such cases.

\bibliographystyle{IEEEtran}

\bibliography{References}

\begin{thebibliography}{10}
\providecommand{\url}[1]{#1}
\csname url@samestyle\endcsname
\providecommand{\newblock}{\relax}
\providecommand{\bibinfo}[2]{#2}
\providecommand{\BIBentrySTDinterwordspacing}{\spaceskip=0pt\relax}
\providecommand{\BIBentryALTinterwordstretchfactor}{4}
\providecommand{\BIBentryALTinterwordspacing}{\spaceskip=\fontdimen2\font plus
\BIBentryALTinterwordstretchfactor\fontdimen3\font minus \fontdimen4\font\relax}
\providecommand{\BIBforeignlanguage}[2]{{%
\expandafter\ifx\csname l@#1\endcsname\relax
\typeout{** WARNING: IEEEtran.bst: No hyphenation pattern has been}%
\typeout{** loaded for the language `#1'. Using the pattern for}%
\typeout{** the default language instead.}%
\else
\language=\csname l@#1\endcsname
\fi
#2}}
\providecommand{\BIBdecl}{\relax}
\BIBdecl

\bibitem{3GPP38827}
{3GPP Technical Specification Group Radio Access Network. Technical report TR 38.827 V16.8.0}, ``{Study on radiated metrics and test methodology for the verification of multi-antenna reception performance of NR User Equipment (UE); (Release 16).}'' Sept 2022.

\bibitem{5GAA_VATM}
\BIBentryALTinterwordspacing
{5GAA}. (2021, Aug) {Vehicular Antenna Test Methodology}. [Online]. Available: \url{https://5gaa.org/content/uploads/2021/08/5GAA_TR_Vehicular_Antenna_Test_Methodology.pdf}
\BIBentrySTDinterwordspacing

\bibitem{5G_Testing_Survey}
P.~Zhang, X.~Yang, J.~Chen, and Y.~Huang, ``{A survey of testing for 5G: Solutions, opportunities, and challenges},'' \emph{China Communications}, vol.~16, no.~1, pp. 69--85, 2019.

\bibitem{STWC}
A.~Paulraj, R.~Nabar, and D.~Gore, \emph{{Introduction to Space-Time Wireless Communication}}.\hskip 1em plus 0.5em minus 0.4em\relax Cambridge, U.K.: Cambridge Univ. Press, 2003.

\bibitem{Rician_Dist_Original}
S.~O. Rice, ``{Mathematical analysis of random noise},'' \emph{The Bell System Technical Journal}, vol.~23, no.~3, pp. 282--332, 1944.

\bibitem{mmWavejust}
T.~S. Rappaport, S.~Sun, R.~Mayzus, H.~Zhao, Y.~Azar, K.~Wang, G.~N. Wong, J.~K. Schulz, M.~Samimi, and F.~Gutierrez, ``{Millimeter Wave Mobile Communications for 5G Cellular: It Will Work!}'' \emph{IEEE Access}, vol.~1, pp. 335--349, 2013.

\bibitem{5Gchannels_survey}
C.-X. Wang, J.~Bian, J.~Sun, W.~Zhang, and M.~Zhang, ``{A Survey of 5G Channel Measurements and Models},'' \emph{IEEE Communications Surveys \& Tutorials}, vol.~20, no.~4, pp. 3142--3168, 2018.

\bibitem{LTEBOOK}
M.~Andersson, A.~Wolfgang, C.~Orlenius, and J.~Carlsson, ``{Measuring performance of 3GPP LTE terminals and small base stations in reverberation chambers},'' in \emph{Long Term Evolution}.\hskip 1em plus 0.5em minus 0.4em\relax Auerbach Publications, 2016, pp. 427--472.

\bibitem{RandomLOS}
P.-S. Kildal, A.~A. Glazunov, J.~Carlsson, and A.~Majidzadeh, ``{Cost-effective measurement setups for testing wireless communication to vehicles in reverberation chambers and anechoic chambers},'' in \emph{2014 IEEE Conference on Antenna Measurements \& Applications (CAMA)}, 2014, pp. 1--4.

\bibitem{Kildal_hyp}
P.-S. Kildal and J.~Carlsson, ``{New approach to OTA testing: RIMP and pure-LOS reference environments \& a hypothesis},'' in \emph{{2013 7th European Conference on Antennas and Propagation (EuCAP)}}, 2013, pp. 315--318.

\bibitem{BluetestCATR}
J.~Kvarnstrand, P.~Svedjenäs, E.~Silfverswärd, and H.~Helmius, ``{Integrating LoS and RIMP Measurements in a Single Test Environment},'' in \emph{2021 15th European Conference on Antennas and Propagation (EuCAP)}, 2021, pp. 1--5.

\bibitem{EuCAP2024_ours}
A.~A. Ruiz, S.~Hosseinzadegan, J.~Kvarnstrand, K.~Arvidsson, and A.~A. Glazunov, ``{K-factor Evaluation in a Hybrid Reverberation Chamber plus CATR OTA Testing Setup},'' in \emph{2024 18th European Conference on Antennas and Propagation (EuCAP)}, 2024.

\bibitem{KFLit1}
C.~M.~J. Wang, K.~A. Remley, A.~T. Kirk, R.~J. Pirkl, C.~L. Holloway, D.~F. Williams, and P.~D. Hale, ``{Parameter Estimation and Uncertainty Evaluation in a Low Rician K-Factor Reverberation-Chamber Environment},'' \emph{IEEE Transactions on Electromagnetic Compatibility}, vol.~56, no.~5, pp. 1002--1012, 2014.

\bibitem{Kildal_RC_KF_formula}
P.-S. Kildal, X.~Chen, C.~Orlenius, M.~Franzen, and C.~S.~L. Patane, ``{Characterization of Reverberation Chambers for OTA Measurements of Wireless Devices: Physical Formulations of Channel Matrix and New Uncertainty Formula},'' \emph{{IEEE Transactions on Antennas and Propagation}}, vol.~60, no.~8, pp. 3875--3891, 2012.

\bibitem{StirUnstir}
X.~Chen, P.-S. Kildal, and S.-H. Lai, ``{Estimation of Average Rician K-Factor and Average Mode Bandwidth in Loaded Reverberation Chamber},'' \emph{IEEE Antennas and Wireless Propagation Letters}, vol.~10, pp. 1437--1440, 2011.

\bibitem{ZaherVIRCKF}
M.~Z. Mahfouz, R.~Vogt-Ardatjew, A.~B.~J. Kokkeler, and A.~A. Glazunov, ``{Measurement and Estimation Methodology for EMC and OTA Testing in the VIRC},'' \emph{IEEE Transactions on Electromagnetic Compatibility}, vol.~65, no.~1, pp. 3--16, 2023.

\bibitem{UncRIMP}
D.~Senic, K.~A. Remley, C.-M.~J. Wang, D.~F. Williams, C.~L. Holloway, D.~C. Ribeiro, and A.~T. Kirk, ``{Estimating and Reducing Uncertainty in Reverberation-Chamber Characterization at Millimeter-Wave Frequencies},'' \emph{IEEE Transactions on Antennas and Propagation}, vol.~64, no.~7, pp. 3130--3140, 2016.

\bibitem{UncRIMP2}
T.~Jia, Y.~Huang, Q.~Xu, Q.~Hua, and L.~Chen, ``{Average Rician K-Factor Based Analytical Uncertainty Model for Total Radiated Power Measurement in a Reverberation Chamber},'' \emph{IEEE Access}, vol.~8, pp. 198\,078--198\,090, 2020.

\bibitem{Emul_Rician_Andres_RC}
A.~A. Glazunov, S.~Prasad, and P.~Handel, ``{Experimental Characterization of the Propagation Channel Along a Very Large Virtual Array in a Reverberation Chamber},'' in \emph{Progress In Electromagnetics Research B}, vol.~59, 2014, pp. 205--217.

\bibitem{KFEmulRician}
C.~Lemoine, E.~Amador, and P.~Besnier, ``{On the $K$ -Factor Estimation for Rician Channel Simulated in Reverberation Chamber},'' \emph{IEEE Transactions on Antennas and Propagation}, vol.~59, no.~3, pp. 1003--1012, 2011.

\bibitem{MIMORician}
J.~D. Sanchez-Heredia, J.~F. Valenzuela-Valdes, A.~M. Martinez-Gonzalez, and D.~A. Sanchez-Hernandez, ``{Emulation of MIMO Rician-Fading Environments With Mode-Stirred Reverberation Chambers},'' \emph{IEEE Transactions on Antennas and Propagation}, vol.~59, no.~2, pp. 654--660, 2011.

\bibitem{PprocRician}
A.~De~Leo, P.~Russo, and V.~Mariani~Primiani, ``{Emulation of the Rician K-Factor of 5G Propagation in a Source Stirred Reverberation Chamber},'' \emph{Electronics}, vol.~12, no.~1, p.~58, Dec. 2022.

\bibitem{Ra_Ri_RMS}
J.-H. Choi, S.-O. Park, T.-S. Yang, and J.-H. Byun, ``{Generation of Rayleigh/Rician Fading Channels With Variable RMS Delay by Changing Boundary Conditions of the Reverberation Chamber},'' \emph{IEEE Antennas and Wireless Propagation Letters}, vol.~9, pp. 510--513, 2010.

\bibitem{RC_Rician}
C.~L. Holloway, D.~A. Hill, J.~M. Ladbury, P.~F. Wilson, G.~Koepke, and J.~Coder, ``{On the Use of Reverberation Chambers to Simulate a Rician Radio Environment for the Testing of Wireless Devices},'' \emph{IEEE Transactions on Antennas and Propagation}, vol.~54, no.~11, pp. 3167--3177, 2006.

\bibitem{ahmed2023overtheair}
I.~Ahmed, M.~Davy, H.~Prod'homme, P.~Besnier, and P.~del Hougne, ``{Over-the-Air Emulation of Electronically Adjustable Rician MIMO Channels in a Programmable-Metasurface-Stirred Reverberation Chamber},'' 2023.

\bibitem{3GPPRCStandard}
{3GPP. Technical report TR 37.941 V17.0}, ``{Radio Frequency (RF) conformance testing background for radiated Base Station (BS) Requirements (Release 17).}'' Mar 2022.

\bibitem{DRH50}
\BIBentryALTinterwordspacing
{RF SPIN}. (2022) {DRH50 Datasheet}. [Online]. Available: \url{https://www.rfspin.com/wp-content/uploads/2022/04/DRH50-–-RF-SPIN.pdf}
\BIBentrySTDinterwordspacing

\bibitem{CTIAprechar}
{CTIA}, ``{01.73 Supporting Procedures.}'' Nov 2023.

\bibitem{CATR}
\BIBentryALTinterwordspacing
{Bluetest}. (2020) {5G OTA DEVICE TESTING IN THE RTS65}. [Online]. Available: \url{https://www.bluetest.se/files/5G\_RevA.pdf}
\BIBentrySTDinterwordspacing

\bibitem{CorrMatrix}
K.~A. Remley, S.~Catteau, A.~Hussain, C.~L. Nogueira, M.~Kristoffersen, J.~Kvarnstrand, B.~Horrocks, J.~Fridén, R.~D. Horansky, and D.~F. Williams, ``{Practical Correlation-Matrix Approaches for Standardized Testing of Wireless Devices in Reverberation Chambers},'' \emph{IEEE Open Journal of Antennas and Propagation}, vol.~4, pp. 408--426, 2023.

\bibitem{CorrMat2}
R.~J. Pirkl, K.~A. Remley, and C.~S.~L. Patane, ``{Reverberation Chamber Measurement Correlation},'' \emph{IEEE Transactions on Electromagnetic Compatibility}, vol.~54, no.~3, pp. 533--545, 2012.

\bibitem{IEC61000421}
{International Electrotechnical Commission, Electromangetic Compatability (EMC)–Part 4–21}, ``{Testing and Measurement Techniques–Reverberation Chamber Test Methods, Standard IEC 61000–4-21},'' Jan 2011.

\bibitem{KFestim_negatives}
C.~M.~J. Wang, K.~A. Remley, A.~T. Kirk, R.~J. Pirkl, C.~L. Holloway, D.~F. Williams, and P.~D. Hale, ``{Parameter Estimation and Uncertainty Evaluation in a Low Rician K-Factor Reverberation-Chamber Environment},'' \emph{IEEE Transactions on Electromagnetic Compatibility}, vol.~56, no.~5, pp. 1002--1012, 2014.

\bibitem{Chisq}
W.~Rolke and C.~Gutierrez~Gongora, ``{A chi-square goodness-of-fit test for continuous distributions against a known alternative},'' \emph{Computational Statistics}, vol.~36, no.~3, pp. 1885--1900, May 2021.

\bibitem{ADGoF}
C.~Lemoine, P.~Besnier, and M.~Drissi, ``{Investigation of Reverberation Chamber Measurements Through High-Power Goodness-of-Fit Tests},'' \emph{IEEE Transactions on Electromagnetic Compatibility}, vol.~49, no.~4, pp. 745--755, 2007.

\bibitem{ADStephens}
M.~Stephens, ``{EDF Statistics for Goodness of Fit and Some Comparisons},'' \emph{Journal of the American Statistical Association}, vol.~69, no. 347, pp. 730--737, 1974.

\bibitem{BootstrapGoF}
{M.A. Stephens}, ``{Bootstrap Based Goodness-Of-Fit-Tests},'' \emph{Metrika}, vol.~40, pp. 243--256, 1993.

\bibitem{Hill_eq}
D.~Hill, M.~Ma, A.~Ondrejka, B.~Riddle, M.~Crawford, and R.~Johnk, ``{Aperture excitation of electrically large, lossy cavities},'' \emph{IEEE Transactions on Electromagnetic Compatibility}, vol.~36, no.~3, pp. 169--178, 1994.

\end{thebibliography}

\end{document}